\documentclass[compsoc,conference,a4paper,10pt,times]{IEEEtran}
\IEEEoverridecommandlockouts
\usepackage{cite}

\usepackage{amsmath,amssymb,amsfonts}

\usepackage{mathtools}
\DeclarePairedDelimiter{\norm}{\lVert}{\rVert}

\usepackage{times}
\usepackage{amsmath, amssymb}
\usepackage{graphicx}
\usepackage{algpseudocode}

\usepackage{cleveref}

\usepackage{color}
\usepackage{algorithm}
\usepackage{todonotes}
\usepackage{dsfont}

\usepackage{graphicx}
\usepackage{textcomp}
\usepackage{bmpsize}
\usepackage{xcolor}
\usepackage{lipsum}

\def\BibTeX{{\rm B\kern-.05em{\sc i\kern-.025em b}\kern-.08em
    T\kern-.1667em\lower.7ex\hbox{E}\kern-.125emX}}
    
\usepackage{tikz}
\graphicspath{ {figs/} }
\usepackage{subcaption}
\usepackage[labelfont=bf]{caption}
\usepackage{fmtcount} 
\usepackage{multirow}
\usepackage{multicol}
\usepackage{booktabs}

\usepackage{array}
\newcolumntype{C}[1]{>{\centering\arraybackslash}m{#1}}
\newcolumntype{L}{>{\centering\arraybackslash}m{3cm}}
\usepackage{color}

\renewcommand{\paragraph}[1]{\vspace{0.04in}\noindent{\bf{#1}.}} 
    
\begin{document}
\pagenumbering{gobble} 

\title{aaeCAPTCHA: The Design and Implementation of Audio Adversarial CAPTCHA}

\author{\IEEEauthorblockN{Md Imran Hossen}
\IEEEauthorblockA{
\textit{University of Louisiana at Lafayette}\\
md-imran.hossen1@louisiana.edu}
\and
\IEEEauthorblockN{Xiali Hei}
\IEEEauthorblockA{
\textit{University of Louisiana at Lafayette}\\
xiali.hei@louisiana.edu}
}

\maketitle


\thispagestyle{plain}
\pagestyle{plain}

\begin{abstract}
CAPTCHAs are designed to prevent malicious bot programs from abusing websites. Most online service providers deploy audio CAPTCHAs as an alternative to text and image CAPTCHAs for visually impaired users. However, prior research investigating the security of audio CAPTCHAs found them highly vulnerable to automated attacks using Automatic Speech Recognition (ASR) systems. To improve the robustness of audio CAPTCHAs against automated abuses, we present the design and implementation of an audio adversarial CAPTCHA (aaeCAPTCHA) system in this paper. The aaeCAPTCHA system exploits audio adversarial examples as CAPTCHAs to prevent the ASR systems from automatically solving them.
Furthermore, we conducted a rigorous security evaluation of our new audio CAPTCHA design against five state-of-the-art DNN-based ASR systems and three commercial Speech-to-Text (STT) services. Our experimental evaluations demonstrate that aaeCAPTCHA is highly secure against these speech recognition technologies, even when the attacker has complete knowledge of the current attacks against audio adversarial examples. We also conducted a usability evaluation of the proof-of-concept implementation of the aaeCAPTCHA scheme. Our results show that it achieves high robustness at a moderate usability cost compared to normal audio CAPTCHAs. Finally, our extensive analysis highlights that aaeCAPTCHA can significantly enhance the security and robustness of traditional audio CAPTCHA systems while maintaining similar usability.



\end{abstract}

\begin{IEEEkeywords}
CAPTCHA, security, audio adversarial CAPTCHA, ASR system, Speech-to-Text service
\end{IEEEkeywords}

\section{Introduction}
CAPTCHAs (Completely Automated Public Turing tests to tell Computers and Humans Apart) are computer-generated and graded tests to distinguish humans from automated bot programs \cite{von2004telling}. Such tests are designed to be easy and intuitive for humans while remaining extremely difficult for current computer programs.  
Millions of websites use CAPTCHAs to safeguard online services from automated abuse. Thus, CAPTCHA is of immense interest to researchers and hackers alike, and intelligent attacks usually lead to better CAPTCHA designs.  

Both text and image CAPTCHAs have been extensively investigated in terms of security. Most earlier CAPTCHAs are text-based. Research works show that the current machine learning algorithms can easily solve the underlying text recognition problems in most text CAPTCHAs automatically and with high accuracy \cite{mori2003recognizing, chellapilla2005using, yan2007breaking, yan2008low, bursztein2011text, bursztein2014end, yan2016simple, ye2018yet}. As text CAPTCHAs become increasingly vulnerable to automated solvers, their popularity has diminished in recent years. 

Image CAPTCHA is currently the most popular CAPTCHA scheme on the Internet. The majority of image CAPTCHAs are based on image classification and object recognition problems. The security guarantee for these problems is based on the presumed difficulty of solving such problems using a computer program automatically. However, recently, deep learning has achieved impressive results in complex image recognition tasks. Previous works developed machine learning, especially deep learning-based, attacks that can break popular image captcha designs with high accuracy \cite{golle2008machine, zhu2010attacks, sivakorn2016robot, weng2019towards, hossen2020object, hossen2021low}. As a result, researchers and CAPTCHA designers are constantly investigating methods to harden the image CAPTCHA designs to resist automated solvers. 

Most online service providers also deploy audio CAPTCHAs as an alternative to text and image CAPTCHAs for visually impaired users. Solving an audio CAPTCHA challenge is sufficient to bypass the reverse Turing test to verify that a web user is a human and not a computer. As a result, the security of audio CAPTCHA is critical for defending the internet against automated bot programs.

While the security of text and image CAPTCHAs has been well studied, the security of audio CAPTCHAs is often overlooked. Nonetheless, a few research works that have examined the security of various real-world audio CAPTCHA systems found that these systems are highly vulnerable to automatic speech recognition (ASR) systems and other deep learning-based attacks \cite{tam2008breaking, bursztein2009decaptcha, bock2017uncaptcha, meutzner2014using,solanki2017cyber}. For example, Bock \textit{et al.} \cite{bock2017uncaptcha} proposed a low-resource attack against reCAPTCHA's auditory challenges using online speech-to-text (STT) APIs. Their attack was able to break the audio reCAPTCHA system with 85.15\% accuracy. Similarly, Solanki \textit{et al.} \cite{solanki2017cyber} examined the security of various online audio CAPTCHA schemes and devised attacks using off-the-shelf STT services. The authors showed that their attack could break all of the audio CAPTCHA systems they analyzed.

Deep neural network (DNN)-based automatic speech recognition (ASR) systems have seen impressive progress in recent years. However, prior research has demonstrated that DNNs are vulnerable to adversarial examples \cite{goodfellow2014explaining,papernot2016limitations, carlini2017towards}. While previously extensively studied in the image domain, recent attacks on ASR systems have highlighted the feasibility of crafting adversarial examples in the audio domain \cite{carlini2018audio, qin2019imperceptible, schonherr2018adversarial}. An audio adversarial example can lead the victim ASR model to make a significant transcription error or cause the original audio input to be transcribed to a target phrase desired by the adversary.

At the same time, efforts have been made to mitigate audio adversarial examples \cite{yang2018characterizing, kwon2019poster, hussain2021waveguard}. Most recently studied countermeasures against audio adversarial examples are based on input transformation or audio preprocessing techniques such as quantization, signal smoothing, filtering, and audio compression. Audio adversarial examples are an open research subject. As a result, most methods for removing or detecting audio adversarial examples are still in the early stages of development. Furthermore, these techniques have been found to be broken under an adaptive attack setting, where the attacker has knowledge of the employed defense techniques. In summary, audio adversarial examples remain a potent threat to the ASR systems.

Based on the above observations, we propose to exploit the vulnerability of ASR systems to audio adversarial examples as a ``defense'' to develop robust audio CAPTCHAs. Audio CAPTCHA challenges in such a CAPTCHA scheme should be easy for humans to solve to maintain good usability. Yet, the ASR systems should be forced to make significant errors while automatically transcribing them. This way, an audio CAPTCHA system harnessing audio adversarial examples can significantly improve the robustness of traditional audio CAPTCHA systems against emerging machine learning attacks. It is worth noting that previous research has proposed the design of text and image adversarial CAPTCHAs \cite{osadchy2017no,shi2021adversarial}. However, the design of the audio adversarial CAPTCHA has not been studied. While some prior research suggests using audio adversarial examples for CAPTCHAs \cite{abdullah2019hear}, the comprehensive security and usability evaluation of such an audio CAPTCHA system remain an unexplored area of research.

In this paper, we present the design and implementation of an audio adversarial CAPTCHA system called \texttt{aaeCAPTCHA}. The basic design idea is that instead of asking users to solve a speech recognition problem involving normal audio, we ask them to solve a problem involving audio with adversarial perturbation. \texttt{aaeCAPTCHA} generates audio adversarial CAPTCHAs on top of current adversarial example generation techniques. Furthermore, we conducted a rigorous security evaluation of our new audio CAPTCHA scheme against five state-of-the-art DNN-based ASR systems and three commercial Speech-to-Text (STT) services. Our experimental results show that while ASR systems and STTs are highly accurate at correctly transcribing normal audios, they make significant transcription errors while transcribing \texttt{aaeCAPTCHA} challenges.

Moreover, we investigated the robustness of \texttt{aaeCAPTCHA} under an adaptive security setting where the attacker has complete knowledge of the current countermeasures against audio adversarial examples. Specifically, we analyzed five audio preprocessing techniques, including quantization, down-sampling, filtering, and audio compression, to break \texttt{aaeCAPTCHA}. We also examined adversarial training as an attack against \texttt{aaeCAPTCHA}. Our extensive experimental results show that most of those attacks are largely ineffective against our system. Finally, we conducted a usability evaluation of the \texttt{aaeCAPTCHA} scheme. Our results show that it achieves high robustness at a moderate usability cost compared to normal audio CAPTCHAs. Our extensive analysis highlights that \texttt{aaeCAPTCHA} can significantly enhance the security and robustness of traditional audio CAPTCHA systems while maintaining similar usability.

In summary, we made the following contributions in this study:
\begin{itemize}
    \item We presented the design and implementation of an audio adversarial CAPTCHA system called \texttt{aaeCAPTCHA}.
    
    \item We conducted a rigorous security evaluation of our new audio CAPTCHA design against five state-of-the-art DNN-based ASR systems and three commercial Speech-to-Text (STT) services. Our experimental evaluations demonstrate that \texttt{aaeCAPTCHA} is highly secure against these speech recognition technologies.
    
    \item We investigated the robustness of the \texttt{aaeCAPTCHA} system against five audio preprocessing methods aiming to remove adversarial perturbations and recover the original transcriptions in our adaptive security evaluation setting. In addition, we also studied adversarial training to break our system. Our experimental results show that most of these attacks are either ineffective or easily circumvented. Overall, our security analysis showed that \texttt{aaeCAPTCHA} could withstand the most advanced attacks, indicating that it could be used to strengthen the security of traditional audio CAPTCHA systems by rendering automated ASR attacks ineffective to a large extent.
    
    
    \item We carried out a user study to evaluate the usability of the \texttt{aaeCAPTCHA} system, and our results show that it maintains good usability. 
\end{itemize} 

\begin{figure*}[t!]
     \centering
        \includegraphics[height=3cm,width=\textwidth]{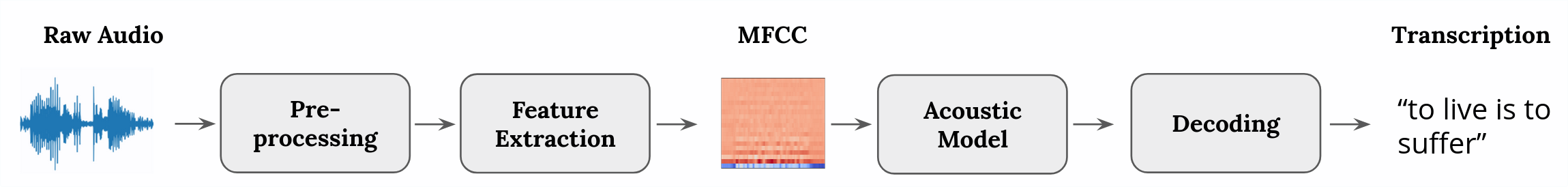}
     \caption{The typical architecture of automatic speech recognition (ASR) systems.}
     \vspace{-3mm} 
    \label{fig:ASR_Arch}
\end{figure*}

\vspace{-3mm} 
\section{Background}
\vspace{-2mm} 
\subsection{Speech Recognition}
\vspace{-2mm}
End-to-end Automatic Speech Recognition (ASR) systems allow machines to turn speech into text automatically and have been widely used in personal assistants and other human-machine interactive devices. Preprocessing, feature extraction, acoustic model, and decoding are the four essential components of a typical ASR system, as depicted in Figure \ref{fig:ASR_Arch}.

\paragraph{Preprocessing} Preprocessing removes noise and interference from the raw audio signal by filtering frequencies beyond the human auditory range, yielding a clean signal. This is generally done by using a combination of noise filters and low-pass filters. Noise filters filter out undesirable frequency components from the audio signal that are unrelated to the speech. Various ASR systems have different processes for identifying and removing noise. Furthermore, because the majority of frequencies in human speech fall between 300 and 3,000 Hz \cite{pierce1990acoustics}, using a low-pass filter to discard higher frequencies helps remove unnecessary information from the audio. 

\paragraph{Feature extraction} Unlike neural network models for image recognition that directly use images as inputs, acoustic models are trained on the features extracted from preprocessed audios. Feature extraction algorithms such as Mel-frequency Cepstral Coefficients (MFCC) \cite{muda2010voice}, Linear Predictive Coefficient (LPC) \cite{doi:10.1121/1.1995189}, and Perceptual Linear Predictive (PLP) \cite{Hermansky1990PerceptualLP} are commonly used in modern ASR systems. Among these, MFCC is the most popular feature extraction method used in modern ASR systems. The MFCC is comprised of several steps. First, the input waveform is divided into overlapping frames of a fixed length. Then, each frame is transformed individually using the Discrete Fourier Transform (DFT) to extract the frequency information. Next, the Mel filter is applied to the magnitude of DFT since it is designed to resemble the human ear. The logarithm of powers is then used, as the human ear perceives loudness on a logarithmic scale. Finally, the Discrete Cosine Transform (DCT) function returns the MFCC coefficients from the logarithm of powers.

It should be noted that some modern ASR systems use data-driven learning methods to estimate which features to extract. Specifically, a neural network layer is often trained to learn which features to extract from an audio sample to transcribe the speech effectively.

\paragraph{Acoustic model} In the early evolution of ASR systems, the Hidden Markov Model (HMM) was one of the preferred methods \cite{10.3115/1075434.1075475}. However, deep neural networks (DNNs) \cite{hinton2012deep} and, in particular, recurrent neural networks (RNNs) \cite{graves2013speech, rao2017exploring} have garnered increasing attention in various speech recognition tasks as deep learning has progressed. The Connectionist Temporal Classification (CTC) \cite{10.1145/1143844.1143891} is one of the preferred methods for training sequence-to-sequence neural networks since it solves the problem of finding alignment between input and output sequences. 

\paragraph{Decoding} Finally, the phoneme probabilities are passed into a decoding algorithm that produces the transcription. Decoding techniques such as greedy decoding are fast; however, they often output transcriptions with many grammatical errors. For this reason, when high-quality transcriptions are required, a more advanced method like beam search decoding \cite{graves2012sequence, BoulangerLewandowski2013AudioCR, sutskever2014sequence} with a language model is often employed. 

\vspace{-2mm}
\section{Related Work}
\vspace{-2mm}

\subsection{Adversarial Examples against ASR systems}
\vspace{-2mm}

\paragraph{White-box attacks}
Early research has shown that ASR systems are inherently vulnerable to audio adversarial examples, particularly in white-box settings \cite{alzantot2018did, carlini2018audio, carlini2016hidden, iter2017generating, vaidya2015cocaine}. Adversarial attacks on ASR systems have generally focused on embedding carefully crafted perturbations into speech signals, causing the victim model to transcribe the input audio into a specific malicious phrase, as desired by the adversary \cite{alzantot2018did, carlini2018audio, carlini2016hidden, iter2017generating, vaidya2015cocaine}. Traditional speech recognition models based on HMMs and GMMs \cite{baum1967inequality, baum1970maximization, acero2000hmm, ahadi1997combined} have been successfully attacked in the past \cite{carlini2016hidden, vaidya2015cocaine}.

Carlini \textit{et al.} \cite{carlini2018audio} were the first to design a gradient-based optimization attack against the end-to-end DeepSpeech \cite{hannun2014deep} ASR model with a 100\% targeted success rate.
However,
the crafted adversarial sample fails when played over the air. Kaldi was reportedly successfully attacked by CommanderSong \cite{yuan2018commandersong}, which embeds malicious commands into popular songs. It also launched a limited over-the-air attack that was significantly reliant on recording devices, speakers, and room settings. Yakura \textit{et al.} \cite{yakura2018robust} attempted to create robust over-the-air adversarial samples by using impulse responses to simulate reverberation, with a success rate of roughly 60\%. The perceptibility of the adversarial perturbations was minimized using psychoacoustic hiding to generate imperceptible adversarial samples in \cite{schonherr2018adversarial}. Moreover, to address the imperceptibility of audio attacks, Qin \textit{et al.} \cite{qin2019imperceptible} used the psychoacoustic principle of auditory masking to create effectively imperceptible audio adversarial examples.

\paragraph{Black-box attacks} While white-box audio adversarial attacks have been shown to be effective against open-source ASR systems, their black-box counterparts have made little progress until lately. In \cite{taori2019targeted}, the authors combined a genetic algorithm with a gradient estimate technique employing CTC-loss to attack DeepSpeech. Aside from the relatively low success rate (35\%) even after many queries, this type of black-box attack is not applicable to commercial systems. An evolutionary multi-objective optimization approach to attack two ASR systems in both un-targeted and targeted settings was introduced in \cite{khare2018adversarial}. Moreover, the particle swarm optimization method, an enhanced optimization genetic algorithm, was utilized to attack black-box ASR models in \cite{du2020sirenattack}. Abdullah \textit{et al.} \cite{abdullah2019hear} developed black-box and transferable attacks by attacking the pipeline stages before the acoustic model to force mistranscription in state-of-the-art systems. By changing only a few audio frames, the authors were able to obtain 100\% mistranscription rates against numerous commercial ASR systems, including Google STT, Facebook Wit, and CMU Sphinx. With 98\% of target commands being successful, the recently developed Devil's Whisper \cite{chen2020devil} demonstrated that adversarial commands embedded in music samples and played over-the-air using speakers can attack commercial ASR systems used in popular intelligent voice control (IVC) devices.

\paragraph{Universal adversarial perturbations} Although numerous effective audio adversarial attacks have been devised and demonstrated, most of them are input-dependent, meaning the perturbation generation approach relies on pre-determined, specific input audio. This limitation renders most existing audio adversarial attacks less effective in the case of real-time audio (\textit{e.g.}, streaming audio as input) because the attacker is unlikely to solve the input-dependent optimization problem in a timely manner. To this end, strategies for generating input-agnostic universal adversarial perturbations have been proposed \cite{neekhara2019universal, vadillo2019universal, xie2020enabling, senior2020stop, li2020advpulse}. The authors of \cite{neekhara2019universal} devised an approach to find a single imperceptible universal perturbation that, when introduced to any arbitrary speech signal, causes the victim speech recognition model to mistranscribe it.
This work also showed the transferability of adversarial audio samples across two different ASR systems (based on DeepSpeech and Wavenet), indicating that audio adversarial attacks could be carried out in real-time even when the attacker is unaware of the ASR model parameters. 

\subsection{Countermeasures against Audio Adversarial Examples}
\vspace{-2mm}

Compared to the image domain, only a few studies have proposed countermeasures against adversarial attacks in the audio domain. Previous research on audio adversarial example mitigation relied primarily on input transformations or audio preprocessing methods (\textit{e.g.}, down-sampling \cite{tamura2019novel, yang2018characterizing, hussain2021waveguard}, signal smoothing \cite{guo2020inor, yang2018characterizing, hussain2021waveguard}, audio compression \cite{das2018adagio, rajaratnam2018isolated, zhang2019defending}, adding distortion \cite{kwon2019poster, rajaratnam2018noise, mendes2020defending}) for detecting or removing adversarial perturbations. 

Yang \textit{et al.} \cite{yang2018characterizing} proposed a method to detect audio adversarial examples that exploits the temporal dependency property in audio data by only transcribing a segment of the audio and comparing the transcription to a transcription of the whole audio using word error rate (WER). The detection technique was evaluated on three attack methods targeting state-of-the-art ASR models like Kaldi and DeepSpeech. The authors also assessed the robustness of their defense framework by performing an adaptive attack on it and showing that the method can withstand such attacks. The authors of \cite{yang2018characterizing} also studied various input transformations, such as quantization, local smoothing, and down-sampling, and found that they are ineffective in adaptive attack settings.


Rajaratnam \textit{et al.} \cite{rajaratnam2018isolated} investigated the use of preprocessing techniques such as audio compression, band-pass filtering, and audio panning in both isolated and ensemble algorithms for detecting audio adversarial examples. Although the authors claimed to have a high detection rate against the targeted adversarial attack, they did not evaluate their methods in an adaptive attack setting.

In a more recent study, Hussian \textit{et al.} \cite{hussain2021waveguard} presented WaveGuard to detect audio adversarial samples utilizing various input transformations such as quantization, down-sampling, and filtering. 
WaveGuard was able to detect adversarial examples generated by four recent attacks \cite{carlini2018audio, qin2019imperceptible, neekhara2019universal} with high success rates. The authors reported that adaptive attacks could easily bypass naive input transformation techniques. However, bypassing LPC and MFCC extraction-inversion requires the adversary to add significant distortions ($||\delta||_\infty>2,000$) to adversarial samples. 

As a countermeasure against audio adversarial examples, applying MP3 compression to adversarial samples to remove any signals below the human perceptibility threshold has been proposed in several studies \cite{das2018adagio, rajaratnam2018isolated, zhang2019defending}. However, these methods often lead to a higher WER/CER on benign samples. Additionally, by accounting for MP3 compression during the optimization process, some attacks have been able to create adversarial examples that are robust to MP3 compression \cite{carlini2018audio}. 

Aside from audio transformations and preprocessing-based countermeasures deployed within the pipeline stages and before the acoustic model in ASR systems, measures have been proposed to enhance the robustness of the acoustic neural network model against audio adversarial examples, specifically through adopting the adversarial training method. The concept of adversarial training was first introduced in \cite{szegedy2013intriguing} to defend against image adversarial attacks by incorporating adversarial examples during the model training stage. A few previous studies have also employed the adversarial training technique in the audio domain \cite{sun2018training, sun2019adversarial, wang2019adversarial}.

\vspace{-2mm}
\section{Audio Adversarial Example Generation}
\vspace{-3mm}

\paragraph{Speech recognition}
This paper focuses on speech-to-text ASR systems that take speech as input and output a transcription of its contents. We let $f(\cdot)$ denote the ASR system. Thus, in the case of $f(x) = y$, the speech input to the ASR system, $x$, provides a transcription $y$.

\paragraph{Adversarial examples} An adversarial example $x'$ is created by adding perturbation $\delta$ to the original audio waveform $x$: $ x' = x + \delta $ such that $f(x) \neq f(x')$. An adversary might try to optimize for a specific value of $t$ (\textit{i.e.}, $f(x') = t$). The term ``targeted'' refers to such adversarial examples.

\paragraph{Problem formulation}
Unlike most prior studies \cite{carlini2018audio, qin2019imperceptible}, which have attempted to create targeted and imperceptible audio adversarial examples, this paper aims to generate untargeted adversarial examples to deceive the ASR system into making incorrect transcriptions. As a result, given an audio waveform $x$ and the ground truth transcription $y$, we search for the following \textbf{untargeted} adversarial perturbation $\delta$: 


\vspace{-3mm}
\begin{equation}
    \begin{split}
      f (x+\delta) &\neq f(x) \\
      \text{such that } & \norm{\delta}_\infty \leq \epsilon \\
      & x + \delta \in [-M,M]
    \end{split}
    \label{Eq:adv}
\end{equation}
\vspace{-3mm}

\noindent where  $\|\cdot\|_\infty$ denotes the $L_\infty$ norm, $\epsilon$ is the magnitude of maximum allowed perturbation, $M$ is the maximum representable value for an audio input ($2^{15}$ in our case).

We define an objective function $\ell$ for finding $\delta$ based on Equation \ref{Eq:adv}. Function $\ell$ is divided into two parts: deceiving the model and minimizing adversarial perturbation amplitude. 
We use the ASR model's loss function $\ell_{net}$ to maximize the distance between the ground truth transcription $y$ and the ASR output for the adversarial example $x'=x+\delta$ to deceive the model. As a result, the first part solves:

\vspace{-3mm}
\begin{equation}
    \begin{split}
      &\text{maximize } \ell_{net} (f(x+\delta), y) \\
      &\text{such that } \norm{\delta}_\infty \leq \epsilon
    \end{split}
\end{equation}

\noindent where $\ell_{net}$ represents model's loss function, and $\epsilon$ is the maximum allowed perturbation and $\|\cdot\|_\infty$ denotes the $L_\infty$ norm. We used the CTC-loss as the $\ell_{net}$, which is also used by our target ASR system DeepSpeech. 

The $L_2$ norm of $\delta$ can be minimized to keep a small amplitude. We use constants $c_1$ and $c_2$ to balance the trade-off between the two parts. Therefore, given an audio input $x$, we aim to minimize:

\begin{equation}
    \begin{split}
        \ell(x,\delta,y) = -c_1 \cdot \ell_{net}(f(x+\delta), y) + c_2 \cdot \norm{\delta}_2
    \end{split}
    \label{Eq:loss_func}
\end{equation}


\paragraph{Crafting audio adversarial examples using PGD} To solve the objective function in Equation \ref{Eq:loss_func}, we use the projected gradient descent (PGD) method \cite{madry2017towards}. We calculate the gradients of loss in Equation \ref{Eq:loss_func} with respect to $\delta$ in each iteration. The sign of the gradients is then multiplied by a learning rate parameter $\alpha$, and the result is subtracted from the current $\delta$.
Finally, we clip $\delta$ to ensure that it stays within an $\epsilon$-neighborhood. As a result, finding the perturbation $\delta$ is done by using:

\begin{equation}
    \begin{split}
         &\delta_0 = 0 \\
         &\delta_{t+1} = clip_{\epsilon} (\delta_t - \alpha \cdot sign (\nabla_\delta \ell(x, \delta, y) )
    \end{split}
\end{equation}

\noindent The complete process is shown in Algorithm \ref{alg:audio-ae-pgd}.

\begin{algorithm}[H]
   \caption{Crafting audio adversarial examples using PGD}
   \label{alg:audio-ae-pgd}
   \hspace*{\algorithmicindent} \textbf{Input:} audio waveform $x$, ground-truth transcription $y$, maximum allowed perturbation $\epsilon$, ASR system $f(\cdot)$, hyperparameters $c_1$ and $c_2$, and number of steps $steps$, learning rate in each step $\alpha$ \\
   \hspace*{\algorithmicindent} \textbf{Output:} audio adversarial example $x + \delta$ 
\begin{algorithmic}[1]
    	\State {\bfseries initialize:} $\delta \gets 0$ 
        \For{$t=1$ {\bfseries to} $steps$}
    	      \State $\ell \gets -c_1 \cdot \ell_{net}(f(x+\delta), y) + c_2 \cdot  \norm{\delta}_2$ 
    	      \State $\delta \gets \delta - \alpha \cdot sign (\nabla_\delta \ell)$
    	      \State $\delta \gets clip_{\epsilon}(\delta)$ 
        \EndFor
\end{algorithmic}
\end{algorithm}

Similar to Carlini and Wagner's approach \cite{carlini2018audio}, our audio adversarial example generation method is end-to-end, operating directly on the raw audio waveforms. However, since our target model DeepSpeech uses MFCC features as the ASR input, the MFCC preprocessing transformation needs to be differentiable to realize this method. We used a differentiable implementation of MFCC in TensorFlow from \cite{carlini2018audio} in our implementation.

\paragraph{Crafting audio adversarial examples using FGSM}
 Single-step methods such as the Fast Gradient Sign Method (FGSM) \cite{goodfellow2014explaining} are an efficient way of crafting adversarial examples. We also evaluated the effectiveness of FGSM in the audio domain to generate audio adversarial examples. FGSM performs the one-step update along the direction of the gradient to maximize the loss, and the crafted adversarial samples stay within $L_\infty$ neighbor of the benign samples. Formally, the adversarial example is $x'$ is obtained as follows:

\begin{equation}
x' = x + \epsilon \cdot sign(\nabla_x\ell_{net}(f(x), y)) 
\end{equation}

\section{Experimental Setup} 
\vspace{-3mm} 

\subsection{Evaluation Metrics} \label{sec:eval_metrics}
\vspace{-1mm}

We employ the following standard metrics for evaluation: 

\paragraph{Word Error Rate (WER)} The word error rate (WER) is a widely used metric for computing ASR system performance. The Levenshtein distance \cite{navarro2001guided} algorithm, which calculates the minimum edit distance between two strings, is used to compute the WER. The WER is defined as the minimum edit distance between an ASR output and the reference transcription. It is calculated as follows:
\begin{equation}
\text{WER}=\frac{S+D+I}{N_W},
\end{equation}

\noindent where the number of substitutions, deletions, and insertions between the reference (ground truth) and actual (model's output) transcriptions are $S$, $D$, and $I$, respectively, and $N_W$ is the number of words in the reference transcription. 





\paragraph{Adversarial success rate ($A_a$)} The adversarial success rate ($A_a$) refers to the ratio of adversarial samples that can successfully mislead a given ASR system. That is, it is the ratio of adversarial samples that are mistranscribed by the target ASR system. An adversarial sample is considered mistranscribed by a given ASR system if the ASR output for the sample returns a non-zero WER with respect to the original transcription $y$. It is calculated as follows:

\begin{equation}
   A_a = \frac{N_f}{N_a}, 
\end{equation}\label{Eq:Aa}

\noindent where $N_a$ is the number of audio adversarial examples that we test and $N_f$ is the number of audio adversarial examples that are mistranscribed by the ASR system. We will use this metric to assess the effectiveness of our adversarial audio generation methods. From a CAPTCHA designer's perspective, we want to achieve higher $A_a$ to ensure our system is robust to ASR attacks. 

\paragraph{Success rate of attack (SRoA)} The success rate of attack (SRoA) is defined as the percentage of audio adversarial examples that are correctly transcribed by the ASR system. An adversarial example is considered correctly transcribed by a given ASR system if the ASR output of the adversarial example $x'$ matches the ground truth transcription $y$ of the natural sample $x$ (\textit{i.e.}, the adversarial example fails to fool the ASR system). That is when $\text{WER}(y, f(x')) = 0$. Essentially, the SRoA is computed as follows:

\begin{equation}
\text{SRoA} = 1-A_a, 
\end{equation}

\noindent The higher the SRoA, the stronger the attack. As such, an adversary aiming to break our CAPTCHA design will attempt to obtain a higher SRoA to make the attack more effective. We will use the SRoA metric during security evaluation to quantify the efficacy of different attacks against our \texttt{aaeCAPTCHA} design. 

\paragraph{Signal-to-noise ratio (SNR)} The SNR quantifies the amount of noise $\delta$, added to the original signal $x$, and is measured in decibels (dB). It is calculated via:

\begin{equation}
\text{SNR (dB)} = 10\cdot\log_{10}\frac{P_x}{P_\delta},
\end{equation}

\noindent where $P_x$ and $P_\delta$ are the energies of the original signal and the perturbation. This means that the higher the SNR ratio, the less distortion is caused by the perturbation. 


\vspace{-1mm}
\subsection{Dataset} \label{sec:dataset}
\vspace{-2mm}

We used the LibriSpeech \cite{panayotov2015librispeech} dataset, a corpus of approximately 1,000 hours of 16 kHz English speech derived from audiobooks from the LibriVox project, in our experiment. Specifically, unless otherwise specified, we conducted all our experiments on 500 randomly selected audio samples with similar transcription lengths from the LibriSpeech \texttt{dev-clean} subset. The transcription length for these audios ranges from 6 to 12 words. The average length of a transcription is 8.98 words.

\vspace{-1mm}
\subsection{Models}
\vspace{-2mm}

In this paper, we focus mainly on DNN-based end-to-end ASR models. Unlike traditional speech recognition systems, these models take audio signals as inputs and output transcriptions without relying on laboriously engineered processing pipelines. Such ASR systems have become very popular for their simplicity and state-of-the-art performance on several speech recognition tasks.

\paragraph{DeepSpeech (target model)} DeepSpeech is an open-source ASR system developed by the Mozilla Foundation based on the original Deep Speech research paper \cite{hannun2014deep}. The system extracts MFCC features from the 16 kHz mono audio signal and uses it as input and outputs transcription of the audio sample. As for the acoustic model, DeepSpeech utilizes a neural network architecture with a combination of dense and LSTM layers. In addition, the network uses connectionist temporal classification (CTC) \cite{10.1145/1143844.1143891} as the loss function for training.

\paragraph{DeepSpeech 2 (DS2)} DeepSpeech 2 \cite{amodei2016deep} is an end-to-end deep neural network model for speech recognition. This model has two convolutional layers, five bidirectional RNN layers, and a fully connected layer. The linear spectrogram extracted from the audio input is the feature in use. Like DeepSpeech 1, DeepSpeech 2 is trained by optimizing the CTC loss function.

\paragraph{Jasper} Jasper \cite{li2019jasper}, a deep time-delay neural network (TDNN), is comprised of only 1D convolutions, batch normalization, ReLU, dropout, and residual connections. Jasper is a collection of models, each with a different number of layers. Jasper models are denoted as Jasper BxR. A Jasper BxR model has B blocks and R repetitions of each convolutional layer within a block. 

\paragraph{Wave2Letter+ (W2L+)} Wave2Letter+ is a fully convolution neural network-based acoustic model based on Facebook's original Wav2letter, and Wav2letterv2 research works \cite{collobert2016wav2letter, liptchinsky1712letter}. The model has 17 1D-convolutional layers and two fully connected layers. NVIDIA's Wave2Letter+ implementation used in this work employs CTC loss for training instead of the Auto SeGmentation (ASG) used in the original Wav2letter implementation. 

\paragraph{Lingvo} The Lingvo \cite{shen2019lingvo} is an attention-based \cite{bahdanau2014neural} sequence-to-sequence \cite{sutskever2014sequence} model based on the Listen, Attend, and Spell model \cite{chan2016listen}. It feeds filter bank spectra into an encoder made up of a stack of convolutional and LSTM layers and outputs the transcription.
The sequence-to-sequence framework enables the entire model to be trained end-to-end.
Lingvo uses the standard cross-entropy loss for training.

\paragraph{Kaldi} Kaldi \cite{povey2011kaldi} is an open-source toolkit for speech recognition developed at Johns Hopkins University. The Kaldi toolkit supports different feature extraction methods like MFCC and fbank. Kaldi is a DNN-HMM-based ASR system that does not employ an end-to-end strategy for speech recognition. Instead, it uses a more typical HMM representation in the decoding stage, with the DNN predicting the likelihood of all HMM states given the acoustic input signal.

We used the DeepSpeech v0.4.1 model pretrained on the LibriSpeech dataset for generating audio adversarial challenges in our CAPTCHA system. For DeepSpeech 2 (DS2), Jasper, and Wave2Letter+ (W2L+), we used pretrained models from the NVIDIA OpenSeq2Seq \cite{openseq2seq} implementation. The Lingvo ASR model was collected from Qin \textit{et al.}'s imperceptible ASR attack \cite{qin2019imperceptible}. For Kaldi, we used the Librispeech ASR Chain 1d model\footnote{https://kaldi-asr.org/models/m13}.

\paragraph{Commercial Speech-to-Text (STT) services} We also tested three commercial STT services against audio adversarial examples generated by our method. The STT APIs are Google Cloud Platform (GCP) STT\footnote{https://cloud.google.com/speech-to-text}, IBM Watson STT\footnote{https://www.ibm.com/cloud/watson-speech-to-text}, and Wit.ai STT\footnote{https://wit.ai/}.

\begin{figure*}[t!]
     \centering
         \includegraphics[height=4.5cm, width=\textwidth]{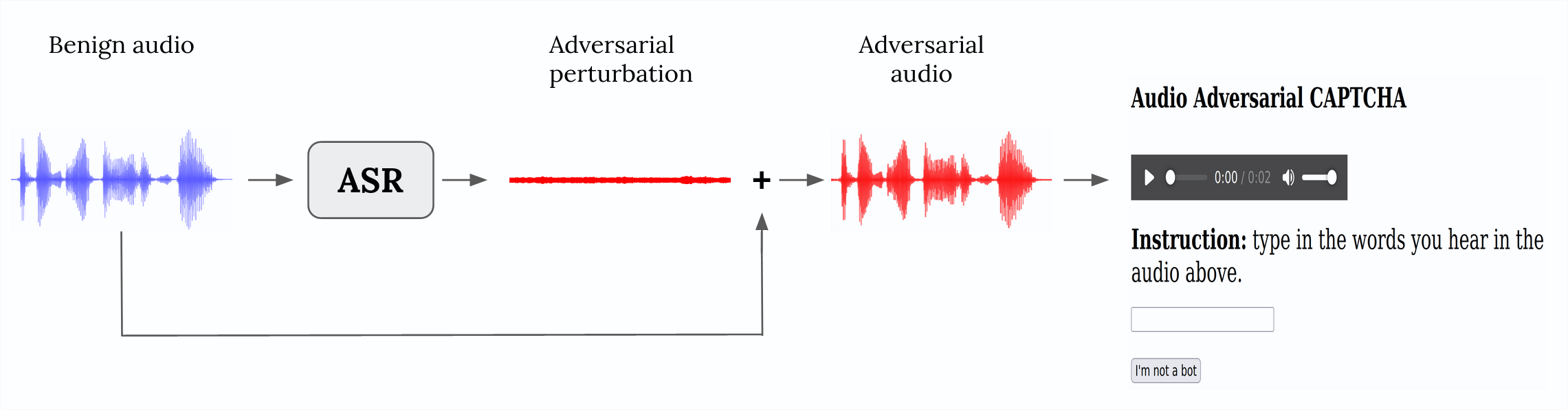}
    
     \caption{High-level system overview of the audio adversarial CAPTCHA (\texttt{aaeCAPTCHA}) system.} 
    \label{fig:audio_adv_captcha_design}
\end{figure*}

\vspace{-2mm} 
\section{Evaluation of Audio Adversarial Example Generation Methods}\label{Sec:Eval_AE_Gen_Methods}
\vspace{-2mm}

All the evaluations in this section were conducted on the DeepSpeech ASR system. Specifically, we used the DeepSpeech v0.4.1 model pretrained on the LibriSpeech dataset.

\begin{table}[tp!]
\centering
\begin{tabular}{|c|c|c|c|c|} 
 \hline
 Metric & $\epsilon:250$ & $\epsilon:300$ &$\epsilon:350$  & $\epsilon:400$  \\ 
 \hline
 WER (\%)    & 55.74 & 59.77 & 63.46 & 67.58 \\ 
 \hline
 $A_a (\%)$  & 94.80 & 96.00 & 97.20 & 98.40 \\
 \hline
 SNR (dB)    & 16.95 & 15.37 & 14.03 & 12.87 \\ 
 \hline
\end{tabular}
\caption{Performance of the FGSM audio adversarial example generation method under various maximum allowed perturbation $\epsilon$ values.}
\vspace{-3mm} 
\label{tab:perf_fgsm}
\end{table}

\vspace{-1mm}
\subsection{FGSM}
\vspace{-2mm}

We investigated the efficacy of the FGSM-based audio adversarial example generation method using various maximum allowed perturbation $\epsilon$ values. During our preliminary analysis, we discovered that an $\epsilon$ value of less than 200 usually does not cause higher transcription errors and thus results in a lower adversarial success rate $A_a$. As a result, we only report our experimental results for $\epsilon=250$ to $\epsilon=400$, with a step size of 50.
Table \ref{tab:perf_fgsm} reports the WER, adversarial success rate $A_a$, and mean signal-to-noise ratio (SNR) on the tested samples. As expected, a higher value of $\epsilon$ induces more transcription errors and improves the adversarial success rate. However, even after increasing the maximum allowed perturbation $\epsilon$ to 400, we still could not achieve an adversarial success rate of 100\%. Furthermore, Table \ref{tab:perf_fgsm} shows that higher $\epsilon$ values appear to deteriorate the mean SNR significantly. As such, while fast, the FGSM method is likely to be less practical for generating adversarial audios in a real-world CAPTCHA system.



\begin{table}[tp!]
\centering
\scriptsize
\begin{tabular}{|C{1.5cm}|C{1cm}|C{1cm}|C{1.5cm}|} 
 \hline
\multirow{2}{4em}{Parameter} & \multicolumn{2}{|c|}{Range} & Step size \\
              & Start & Stop   & \\ 
 \hline
  $\epsilon$  & 250 & 400 & 50 \\ 
  \hline
  $steps$     & 20  & 100 & 10 \\
  \hline
  $\alpha$    & 20  & 100 & 10 \\ 
 \hline
\end{tabular}
\caption{Experimental setup for PGD hyperparameter tuning.} \vspace{-2mm}
\label{tab:pgd_hyper_tuning_setting}
\end{table}

\begin{table*}[tp!]
\centering
\scriptsize
\begin{tabular}{ |C{1.2cm}|C{.8cm}|C{.8cm}|C{.8cm}|C{.8cm}|C{.8cm}|C{.8cm}|C{.8cm}|C{.8cm}|C{.8cm}|C{.8cm}|C{.8cm}|C{.8cm}|  }
 \hline
 \multirow{2}{4em}{Metric} & \multicolumn{3}{|c|}{$\epsilon:250$} & \multicolumn{3}{|c|}{$\epsilon:300$} & \multicolumn{3}{|c|}{$\epsilon:350$} & \multicolumn{3}{|c|}{$\epsilon:400$} \\
     & Mean & Med & Std & Mean & Med & Std & Mean & Med & Std & Mean & Med & Std \\
      \hline
  WER (\%)        & 111.82 & 112.27 & 2.37  & 112.26 & 112.94 & 2.73  & 112.53 & 112.84 & 2.78  & 112.18 & 112.94 & 2.85 \\
  \hline
  $A_{a}$ (\%)       & 100.00   & 100.00   & 0.00  & 100.00   & 100.00   & 0.00  & 100.00   & 100.00   & 0.00  & 100.00   & 100.00   & 0.00 \\
  \hline
  SNR (dB)        & 17.53  & 17.34  & 0.52  & 16.16  & 15.98  & 0.63  & 15.04  & 14.87  & 0.71  & 14.10  & 13.94  & 0.76 \\
  \hline
  $||\delta||_1$ & 227.40 & 233.62 & 18.88 & 263.51 & 270.12 & 26.05 & 296.80 & 302.89 & 32.89 & 327.78 & 332.99 & 39.09 \\
 \hline
\end{tabular}
\caption{Performance of the PGD audio adversarial example generation method under various maximum allowed perturbation $\epsilon$ values. \textbf{Med:} Median, \textbf{Std:} Standard Deviation.}
\label{tab:pgd_hp_tuning_results}
\end{table*}

\vspace{-2mm}
\subsection{PGD}
\vspace{-2mm}

Projection gradient descent (PGD) is one of the most powerful white-box methods for creating adversarial examples. We analyzed the effectiveness of PGD under different settings. We used the following global parameters in Algorithm \ref{alg:audio-ae-pgd} throughout this paper when crafting PGD-based adversarial examples: $c_1=1$, and $c_2=0$. We set $c_2=0$ since we did not emphasize much on minimizing the $L_2$ norm of $\delta$ and stressed more on deceiving the ASR system.

\paragraph{PGD hyperparameters tuning} We conducted an extensive experiment to search for the optimal hyperparameters of the PGD algorithm. After setting $c_1=1$ and $c_2=0$, the remaining tunable parameters in Algorithm \ref{alg:audio-ae-pgd} are maximum perturbation magnitude $\epsilon$, the number of PGD steps $steps$, and learning rate in each step $\alpha$. Table \ref{tab:pgd_hyper_tuning_setting} shows the experimental setup for this experiment. We only report results for $\epsilon=250$ to $\epsilon=400$, as with FGSM. While a perturbation budget $\epsilon$ as low as 200 often results in a 100\% adversarial success rate, the WER usually remains below or around 80\%. For this reason, we began our search for $\epsilon$ at 250.

The results of this experiment are shown in Table \ref{tab:pgd_hp_tuning_results}. According to our experimental setup in Table \ref{tab:pgd_hyper_tuning_setting}, we have $9*9=81$ combinations of $steps$ and $\alpha$ to choose from for each perturbation budget $\epsilon$. Since we used 500 audios for the experiment, we generated 40,500 adversarial audios for each $\epsilon$, totaling 162,000 adversarial audios for four different $\epsilon$ values. The reported results are computed across all these audios. In addition to the metrics reported in Section \ref{sec:eval_metrics}, we also report $||\delta||_1$, which is the $L_1$ norm of perturbation $\delta$ and is calculated by summing the absolute values of the $\delta$. The $||\delta||_1$ helps estimate the amount of distortion introduced to audio by the PGD algorithm.

We achieved 100\% adversarial success rates for any combination of $\epsilon$, $steps$, and $\alpha$, as shown in Table \ref{tab:pgd_hp_tuning_results}. Higher $steps$ and $\alpha$ values for any given $\epsilon$ usually result in higher word error rates. However, after some point, the WER gets saturated and does not increase sharply anymore. Increasing the $steps$, and $\alpha$ also increases $||\delta||_1$, and negatively impacts the SNR. Since SNR is directly related to the quality of adversarial samples, a lower SNR value would degrade the overall usability of these audios. Therefore, the CAPTCHA designer must consider all these factors while selecting optimal hyperparameters for the PGD algorithm in a practical audio adversarial CAPTCHA deployment. Finally, based on the experimental results in Table \ref{tab:pgd_hp_tuning_results}, we conclude that PGD allows us to create more fine-grained adversarial audios when compared to FGSM.

\vspace{-2mm}
\section{Design and Implementation of aaeCaptcha}
\vspace{-3mm}

\paragraph{Overview}
We designed and developed a proof-of-concept implementation of the \texttt{aaeCAPTCHA} system. Figure \ref{fig:audio_adv_captcha_design} depicts the overview of the CAPTCHA system. The system consists of two major components: 1) the audio adversarial CAPTCHA generator and 2) the client interface. The audio adversarial CAPTCHA generator module uses an audio database and computes adversarial perturbations against the target ASR system. The client interface is where the users solve the audio adversarial CAPTCHAs. First, it asks users to recognize the speech in adversarial audios to verify that they are humans and not bots. Then, the users prove their humanness by correctly transcribing the audio.
Because audio adversarial CAPTCHAs are designed to deceive ASR systems, most of them will be mistranscribed by computer programs using such technologies. Note that the primary purpose of this study is not to develop the most usable CAPTCHA scheme. Rather, audio CAPTCHAs generated by \texttt{aaeCAPTCHA} should be intelligible to humans while being robust to ASR attacks.

\subsection{Requirements for Audio Adversarial CAPTCHA Generation}
\vspace{-2mm}

An algorithm should ideally meet the following requirements for constructing the adversarial perturbation used in audio CAPTCHA generation:

 \paragraph{Adversarial} The added perturbation should successfully deceive the target ASR with high adversarial success rates.

 \paragraph{Transferable} In addition, generated adversarial audios should be highly transferable to major ASR models since the adversary could use any ASR model or speech-to-text (STT) service to attack the system. 

 \paragraph{Robust} The adversarial perturbation should be difficult to remove or mitigate by any audio preprocessing techniques or other types of attacks.

 \paragraph{Intelligible} The generated adversarial audios should maintain good intelligibility to humans, but imperceptibility or stealthiness is not one of our system's requirements. Increasing the perturbation may prevent the attacker's ASR model from producing correct transcriptions, but the resulting audio CAPTCHAs are likely to be less comprehensible to humans. In an ideal world, audio adversarial CAPTCHAs would have similar recognition accuracy as unperturbed or regular audios for humans. 

 \paragraph{Efficient} The algorithm should be fast and computationally efficient to be useful in a real-world CAPTCHA system.

It is worth noting that previous methods for generating audio adversarial examples are not suitable for audio CAPTCHA systems. Most recent audio adversarial attacks, for example, focus on crafting imperceptible and targeted adversarial samples. These methods are mostly slow and inefficient. For instance, on a single NVIDIA 1080Ti GPU, Carlini and Wagner's \cite{carlini2018audio} optimization-based approach takes around an hour to generate one adversarial example. Furthermore, these algorithms typically construct adversarial samples targeted at a specific ASR system. As a result, the generated adversarial samples are not transferable to other ASR systems. However, high transferability is one of the most critical requirements of our \texttt{aaeCAPTCHA} scheme.


\begin{table}[tp!]
    \centering
    \begin{tabular}{|C{1.7cm}|C{1cm}|C{1cm}|C{1cm}|C{1cm}|}
    \hline
    Parameters & WER (\%) & $A_a$ (\%) & SNR (dB) & $||\delta||_1$ \\
    \hline
    $\epsilon:350$, $steps:50$, $\alpha:40$ & 111.88 & 100.00 & 15.89 & 256.61 \\
    \hline
 \end{tabular}
    \caption{Audio adversarial CAPTCHA generation performance of PGD with selected optimal hyperparameters.} \vspace{-2mm}
    \label{tab:pgd_final_hp_perf}
\end{table}

\subsubsection{The algorithm and its hyperparameter selection}\label{sec:pgd_opt_hp}
Based on the experimental results in Section \ref{Sec:Eval_AE_Gen_Methods}, we opted to select PGD for generating audio CAPTCHAs. As discussed earlier, using a higher perturbation budget $\epsilon$ with larger values of $steps$, and $\alpha$ usually results in more robust adversarial audios. However, this, at the same time, degrades the audio quality considerably. Therefore, to balance the trade-off between security and usability, we need to select hyperparameters for the PGD algorithm that satisfy the requirements outlined above. While searching for optimal hyperparameters for PGD, we observed that adversarial samples with a $||\delta||_1$ below 230 are often vulnerable to commercial STT services like GCP STT and Wit.ai STT. Specifically, these services could correctly transcribe such adversarial audios with a success rate of more than 20\%. As such, we then looked for the hyperparameters that could produce adversarial audios robust enough to deceive most ASR models with high adversarial success rates while minimally degrading the SNR to keep the audios intelligible to humans.

Considering all these factors, we found the following PGD hyperparameters to be optimal for \texttt{aaeCAPTCHA} design: $\epsilon = 350$, $steps=50$, and $\alpha=40$. Table \ref{tab:pgd_final_hp_perf} shows the WER, $A_a$, mean SNR, and mean $||\delta||_1$ for audio adversarial CAPTCHAs crafted using these hyperparameters. Section \ref{sec:usability} discusses in-depth the impact of PGD algorithm hyperparameter selection, particularly the maximum allowed perturbation $\epsilon$, on usability.

\paragraph{ASR model and dataset} We employed DeepSpeech as our target ASR model for audio adversarial CAPTCHA generation purposes. We used 500 randomly selected audios with transcription lengths ranging from 6 to 12 words from the LibriSpeech \texttt{dev-clean} subset in our implementation.

\paragraph{Implementation and evaluation platforms} Most of our code was developed using TensorFlow 1.12 with Python 3.6 to ensure compatibility with the DeepSpeech v0.4.1 ASR model's codebase. All of our experiments were carried out on a server equipped with 36 Intel Core i9-10980XE CPUs, 251 GB of memory, and four NVIDIA Quadro RTX 8000 GPUs running Ubuntu 18.04.

\paragraph{Computation time} We assume CAPTCHAs are generated offline and then served to the users. Using a batch size of 50, generating 500 CAPTCHAs using PGD with the optimal hyperparameters described earlier takes 3.67 minutes on average on the above platform.


%
%
%
%

\begin{table*}[tp!]
\centering
\begin{tabular}{ |C{1.2cm}|C{.8cm}|C{.8cm}|C{.8cm}|C{.8cm}|C{.8cm}|C{.8cm}|C{.8cm}|C{.8cm}|C{.8cm}|C{.8cm}|C{.8cm}|C{.8cm}|  }
 \hline
 \multirow{2}{4em}{Metric} & \multicolumn{2}{|c|}{DeepSpeech} & \multicolumn{2}{|c|}{DS2} & \multicolumn{2}{|c|}{Jasper} & \multicolumn{2}{|c|}{W2L+} & \multicolumn{2}{|c|}{Lingvo} & \multicolumn{2}{|c|}{Kaldi} \\
     & Normal & Adv & Normal & Adv & Normal & Adv & Normal & Adv & Normal & Adv & Normal & Adv \\
      \hline
  WER (\%) & 6.63 & 111.88 & 5.65 & 78.65 & 2.49 & 56.32 & 4.83 & 76.40 & 2.98 & 74.64 & 8.32 & 80.45 \\
  \hline
  SRoA (\%) & 66.00& 0.00& 68.60 & 0.40& 84.40   & 4.40 & 71.40 & 0.40 & 82.20 & 0.40 & 59.20 & 0.40 \\
 \hline
\end{tabular}
\caption{Attack performance of state-of-the-art open-source ASR models against \texttt{aaeCAPTCHA}. \textbf{Normal:} Normal Audios (Baseline), \textbf{Adv:} Adversarial Audios.}
\label{tab:transf_opnsrc}
\end{table*}

\begin{table*}[tp!]
\centering
\begin{tabular}{ |C{1.5cm}|C{1cm}|C{1cm}|C{1.5cm}|C{1.5cm}|C{1.5cm}|C{1.5cm}|C{1.5cm}|C{1.5cm}|  }
 \hline
 \multirow{2}{4em}{Metric}  & \multicolumn{2}{|c|}{GCP STT} & \multicolumn{2}{|c|}{IBM Watson STT} & \multicolumn{2}{|c|}{Wit.ai STT} \\
      & Normal & Adv & Normal & Adv & Normal & Adv \\
      \hline
  WER (\%)  &   8.37 & 34.75 & 7.34 & 58.74 & 5.52 & 36.57 \\
  \hline
  SRoA (\%) &  61.60 & 17.60  & 63.80  & 5.60 & 68.00  & 12.20 \\
 \hline
\end{tabular} \vspace{-1mm} 
 \caption{Attack performance of various commercial speech-to-text (STT) services against \texttt{aaeCAPTCHA}. \textbf{Normal:} Normal Audios (Baseline), \textbf{Adv:} Adversarial Audios.}
\label{tab:transf_commer_stt}
\end{table*}

\vspace{-1mm}
\section{Security Evaluation}\label{sec:sec_eval}
\vspace{-2mm}

We begin by defining the attacker model and then comprehensively investigate the most prominent attacks that could be used against \texttt{aaeCAPTCHA}.

\vspace{-2mm}
\subsection{The Attacker Model}
\vspace{-2.5mm}

\paragraph{Knowledge of adversarial example generation and attacks against it} The attacker possesses full knowledge of audio adversarial example generation methods and current attacks aimed at mitigating them. 

\paragraph{Knowledge of the audio adversarial CAPTCHA generation algorithm and its internal parameters} The attacker has complete knowledge of the audio adversarial CAPTCHA generation algorithm, including its internal parameters. In our case, the adversary is fully aware that \texttt{aaeCAPTCHA} constructs adversarial CAPTCHAs using the PGD algorithm with $\epsilon=350$, $steps=50$, and $\alpha=40$. Furthermore, we assume the attacker has full knowledge of the target ASR model's architecture and the parameters used in the \texttt{aaeCAPTCHA} system.
This enables the attacker to train a similar ASR model to attack the CAPTCHA system.

\paragraph{Access to audio adversarial CAPTCHA challenges} The attacker has access to all the generated adversarial CAPTCHAs but not their source. The attacker can build an automated program to probe the client interface to obtain as many audio adversarial CAPTCHA challenges as possible.

\paragraph{No access to the original source audios} The attacker does not have access to the source audios used to generate adversarial samples and the ground truth transcriptions for these audio sources.

\paragraph{Access to other speech recognition tools} The attacker has the ability to use any state-of-the-art DNN-based ASR model or STT API to transcribe audio adversarial CAPTCHAs generated by our system.

\vspace{-1mm}
\subsection{General Security Evaluation}\label{sec:gen_sec_eval}
\vspace{-3mm}

Under our threat model, an attacker is open to using any locally trained or commercial ASR model to break the \texttt{aaeCAPTCHA} system. In such a scenario, the audio adversarial CAPTCHAs generated by our system should exhibit high \textbf{transferability} to prevent such ASR models from automatically solving them.
An adversarial sample is called ``transferable'' if it not only fools the target model A but also deceives an unknown model B with similar or different architecture.
The transferability of adversarial examples has mainly been studied in the image domain. Previous work shows that attacks generalize across models in the image domain. In contrast, the transferability of audio adversarial examples has only seen limited success. This subsection analyzes the transferability of our audio adversarial CAPTCHAs against five state-of-the-art open-source ASR models and three commercial (black-box) speech-to-text (STT) services.
Transcriptions generated by these models for a sample audio waveform with the ground truth transcription ``yes he's mercurial in all his movements," and its adversarial version are shown in Appendix \ref{sec:appndx} (Table \ref{tab:asr_out_sample}). While the models correctly transcribed the benign input with high accuracy, they mistranscribed the adversarial example with high transcription errors. Our experimental findings are detailed below.

\paragraph{Evaluation setup} We used the dataset described in Section \ref{sec:dataset} to generate 500 audio adversarial CAPTCHAs against our target ASR model DeepSpeech using the PGD algorithm with the optimal hyperparameters ($\epsilon=350$, $steps=50$, and $\alpha=40$) discussed in Section \ref{sec:pgd_opt_hp}. We set the baseline as the attack performance of normal audios.


\paragraph{Attack using open-source or locally trained ASR models}
Table \ref{tab:transf_opnsrc} lists the WER and success rate of attack (SRoA) of the five state-of-the-art ASR systems on both normal (benign) and adversarial audio samples. Except for Kaldi, all of the models have a WER of less than 7\% on normal audios. The Jasper ASR obtains the highest SRoA, which is over 84\%. This means that if an audio CAPTCHA system used these normal audios, the Jasper ASR would pass the challenges with an accuracy of more than 84\%, rendering the CAPTCHA system broken completely. Similarly, DeepSpeech 2 (DS2), Wave2Letter+ (W2L+), and Lingvo achieved high success rates of attacks against normal audios, showing that these models are highly accurate at correctly transcribing them.

We can see from the results in Table \ref{tab:transf_opnsrc} that ASR models make significant transcription errors while transcribing our adversarial audios. For example, the Kaldi ASR model produces more than 80\% WER, indicating that it is extremely vulnerable to our adversarial audios. This is interesting because the Kaldi toolkit follows a slightly different approach to speech recognition than the remaining ASR models tested in this work. Similarly, DS2, Jasper, W2L+, and Lingvo models produce high transcription errors on adversarial audios. The SRoA for DS2, W2L+, Lingvo, and Kaldi is all below 0.50\% against these adversarial audios. The Jasper ASR model achieves the highest SRoA, which is only 4.40\%.

\paragraph{Attack using commercial speech-to-text (STT) services}
We also evaluated the security of our audio adversarial CAPTCHA scheme against three popular commercial speech-to-text (STT) services: Google Cloud Platform (GCP) STT, IBM Watson STT, and Wit.ai STT. These cloud-based STT services are proprietary, and the specific architectures of these ASR models are not publicly available. The attack performance of the STTs on both benign and adversarial samples is depicted in Table \ref{tab:transf_commer_stt}. On normal audio samples, STTs perform slightly worse than their free and open-source counterparts described above. This could be because open-source ASR models are specifically trained on the LibriSpeech dataset, a subset of which was also used in this study, and hence outperform commercial STTs.
Despite this, commercial STT APIs can correctly transcribe normal audio CAPTCHAs with a success rate of over 61\%.

Table \ref{tab:transf_commer_stt} shows that the STT services induce considerable WER while transcribing adversarial audios. GCP STT, for example, has a WER of 8.37\% on normal audio but a WER of over 34\% on adversarial audio, which is more than four times higher. Among these three, IBM Watson STT has the lowest SRoA, while GCP STT has the highest SRoA against our system.

While commercial STTs show more robustness against our audio adversarial CAPTCHAs than other state-of-the-art ASR models, their attack success rates are still significantly lower when compared to normal audio CAPTCHAs. Furthermore, speech-to-text APIs are paid services, making them an unprofitable attack method for fraudsters who typically rely on low-cost techniques to conduct large-scale attacks.


\vspace{-1mm}
\subsection{Adaptive Security Evaluation} \label{sec:adative_sec_eval}
\vspace{-2mm}

In Section \ref{sec:gen_sec_eval}, we conducted the security analysis of the \texttt{aaeCAPTCHA} system in a non-adapting setting where the attackers do not have knowledge of existing attacks against adversarial audios. In this subsection, we evaluate the \textbf{robustness} of the CAPTCHA system in an adaptive setting with a more prominent adversary that tries to remove or mitigate audio adversarial perturbations to recover the original transcriptions using state-of-the-art attacks on audio adversarial examples. We investigated the robustness of \texttt{aaeCAPTCHA} against five audio preprocessing attacks, including quantization, local signal smoothing, and down-sampling, as well as robust ASR model training through adversarial training.

Note that, in the adversarial machine learning literature, adversarial examples are considered attacks, and any attempts to subvert such attacks are considered countermeasures or defenses. However, \texttt{aaeCAPTCHA} employs adversarial examples as a ``defense'' mechanism, and existing defenses against adversarial examples are used by an adversary that attempts to remove adversarial perturbations to break our CAPTCHA system. As such, we will refer to the traditional adversarial example defenses as ``attacks'' in the subsequent sections.

In Section \ref{sec:adative_sec_eval:defs_target}, we analyze the robustness of our audio adversarial CAPTCHAs against different audio preprocessing attacks on our target ASR model DeepSpeech. Note that while we report SNR when reporting the results (Tables \ref{tab:def_quant}--\ref{tab:def_compression}), the reported SNR has \textbf{no impact on usability} since it is the adversary, not the CAPTCHA designer, who uses audio preprocessing attacks to transcribe adversarial audios correctly. Next, in Section \ref{sec:adative_sec_eval:defs_all}, we study the impact of these attacks on other ASR systems that the attacker might use to break \texttt{aaeCAPTCHA}. Finally, in Section \ref{sec:adative_sec_eval:AT}, we conduct the adversarial training attack against the \texttt{aaeCAPTCHA} system.

\paragraph{Evaluation setup} We used the exact setup that is used in Section \ref{sec:gen_sec_eval}. In particular, we used the 500 adversarial audios created by PGD with $\epsilon=350$, $steps=50$, and $\alpha=40$ against the DeepSpeech ASR model.

\vspace{-2mm}
\subsubsection{Audio preprocessing attacks against aaeCAPTCHA}\label{sec:adative_sec_eval:defs_target} ~\\
\vspace{-3mm}

\paragraph{Quantization \cite{yang2018characterizing, hussain2021waveguard}}
The quantization-based attacks have been used in several studies to neutralize the effect of audio adversarial perturbations. The adversarial perturbation could be disrupted by rounding the 16-bit signed integer amplitude values to the nearest integer multiple of $q$ because its amplitude is usually small in the input space. 

We tested $q=128,\ 256, \ 512, \ 1024$ for conducting quantization-based audio preprocessing attacks. Table \ref{tab:def_quant} depicts the results. Table \ref{tab:def_quant} shows that quantization is generally ineffective against our adversarial audios because it does not substantially lower the WER. As a result, the success rates of attacks stay at 0\% for all tested values of $q$. However, there is a large reduction in the WER for $q=1024$, but it is still significantly high.

\begin{table}[tp!]
\centering
\begin{tabular}{|c|c|c|c|c|} 
 \hline
 Metric & $q=128$ & $q=256$ &$q=512$  & $q=1024$  \\ 
 \hline
  WER (\%) & 109.51 & 108.87 & 107.46 &\textbf{ 87.69} \\ 
  SRoA (\%) & 0.00 & 0.00 & 0.00 & 0.00 \\
  SNR (dB) & 15.74 & 15.85 & 14.29 & 14.42 \\ 
 \hline
\end{tabular}\vspace{-1mm} 
\caption{Robustness of \texttt{aaeCAPTCHA} under Quantization attack.}\vspace{-3mm} 
\label{tab:def_quant}
\end{table}

\begin{table*}[t!]
\centering
\begin{tabular}{|c|c|c|c|c|c|c|c|} 
 \hline
 Metric & $k=3$ & $k=5$ &$k=7$ & $k=9$ & $k=13$ & $k=15$ & $k=17$  \\ 
 \hline
  WER (\%) & 107.80 & 106.33 & 104.57 & 103.18 & 102.38 & 102.42 & \textbf{100.53}\\ 
  SRoA (\%) & 0.00  & 0.00  & 0.00   & 0.00  & 0.00  & 0.00  & 0.00\\
  SNR (dB) & 14.88 & 12.37 & 10.62  & 9.17 & 6.88 & 6.01 & 5.25 \\ 
 \hline
\end{tabular}
\caption{Robustness of \texttt{aaeCAPTCHA} under Average Smoothing attack.}
\label{tab:def_AS}
\end{table*}

\begin{table*}[t!]
\centering
\begin{tabular}{|c|c|c|c|c|c|c|c|} 
 \hline
 Metric & $k=3$ & $k=5$ &$k=7$ & $k=9$ & $k=13$ & $k=15$ & $k=17$  \\ 
 \hline
  WER (\%)& 106.26 & 100.33 & 91.23 & \textbf{84.95} & 86.87 & 89.32 &  90.50 \\ 
  SRoA (\%)& 0.00 & 0.00  & 0.00  & 0.00  & 0.00  & 0.00 & 0.00\\
  SNR (dB) & 13.99 & 12.42  & 11.16 & 10.18 & 7.92 & 6.90 & 6.01 \\ 
 \hline
\end{tabular}
\caption{Robustness of \texttt{aaeCAPTCHA} under Median Smoothing attack.}
\label{tab:def_MS}
\end{table*}

\begin{table*}[tp!]
\centering
\begin{tabular}{|c|c|c|c|c|c|c|c|c|c|c|} 
 \hline
 Metric & \multicolumn{5}{|c|}{Low-pass Filtering} & \multicolumn{5}{|c|}{Band-pass Filtering} \\ 
   & 1.5 kHz & 2 kHz & 3 kHz & 4 kHz & 6 kHz & 1 kHz & 2kHz & 3 kHz & 4 kHz & 6 kHz \\ 
   \hline
  WER (\%)& \textbf{85.88} & 90.94 & 102.02 & 105.46 & 107.26 & 91.58 & \textbf{89.98} & 101.09 & 104.56 & 106.77 \\
  SRoA (\%)& 0.00 & 0.00  & 0.00   & 0.00   & 0.00   & 0.00  & 0.20  & 0.00   & 0.00   & 0.00   \\
  SNR (dB)& 11.73& 13.16  & 14.63  & 15.51  & 16.03  & 8.61  & 11.78 & 13.09  & 13.84  & 14.08 \\
 \hline
\end{tabular}\vspace{-1mm} 
\caption{Robustness of \texttt{aaeCAPTCHA} under Filtering attack.}\vspace{-3mm} 
\label{tab:def_Filter}
\end{table*}

\paragraph{Local smoothing \cite{yang2018characterizing, guo2020inor}}
To limit adversarial perturbation, we investigated two local smoothing attack techniques: 1) average smoothing and 2) median smoothing. Average smoothing reduces adversarial perturbation by applying mean smoothing to the waveform of the adversarial sample. Precisely, for a sampling point $x_i$, the $k-1$ points before and after it are considered local reference points, and $x_i$ is replaced by the average value of its reference points. Median smoothing is similar to average smoothing, except the voice element $x_i$ is replaced with the median value of its local reference points. 

We tested different values of $k$ for both of these local smoothing attacks. Tables \ref{tab:def_AS} and \ref{tab:def_MS} show the results of our experiment. The average smoothing attack appears ineffective in recovering the original transcriptions from the adversarial samples, while median smoothing with $k=9$ resulted in the highest WER decrease.

\begin{table}[tp!]
\centering
\begin{tabular}{|c|c|c|c|c|} 
 \hline
 Metric & 5.6 kHz & 6.4 kHz& 7.2 kHz  & 8 kHz  \\ 
 \hline
  WER (\%) & \textbf{88.51} & 93.20 & 98.48 & 102.71 \\ 
  SRoA (\%)& 0.00  & 0.00  & 0.00  & 0.00 \\
  SNR (dB) & 13.80 & 14.31 & 14.78 & 15.12 \\ 
 \hline
\end{tabular}
\caption{Robustness of \texttt{aaeCAPTCHA} under Down-sampling attack.} \vspace{-3mm}
\label{tab:def_DS}
\end{table}

\paragraph{Down-sampling \cite{tamura2019novel, yang2018characterizing, hussain2021waveguard}}
An audio waveform is down-sampled to a lower sampling rate during a down-sampling attack. The signal recovery is then used to estimate the waveform at its original sampling rate using interpolation. By discarding samples from an audio waveform during the process, down-sampling helps mitigate the adversarial perturbation.
We examined the impact of down-sampling from the attacker's perspective by down-sampling our original 16 kHz audio inputs to lower sampling rates using various down-sampling rates.

Table \ref{tab:def_DS} lists the WER and SRoA for this experiment. While the SRoA stays at 0\% for all tested down-sampling rates, down-sampling the audios at 5.6 kHz considerably reduces the WER.

\paragraph{Filtering \cite{kwon2019poster, rajaratnam2018noise,eisenhofer2021dompteur, hussain2021waveguard}} 
Filtering is frequently used for noise reduction applications, like removing background noise from a speech signal. Several works \cite{kwon2019poster, rajaratnam2018noise, eisenhofer2021dompteur} exploit filtering to disrupt adversarial perturbation in audio samples. We analyzed the low-pass and band-pass filters for the same purpose. Low-pass filtering works by assuming that human speech has lower frequencies than adversarial perturbations and applying a low-pass filter to remove the high-frequency perturbations. The band-pass filter combines a low-pass filter and a high-pass filter to limit the signal's frequency range by removing frequencies below and above certain thresholds to remove more adversarial perturbations outside the typical human speech frequency range.

We tested different cutoff frequencies ranging from 1.5 kHz to 6 kHz for conducting the low-pass filtering attack. We set the lower-cutoff frequency to 100 Hz for band-pass filtering and varied the upper-cutoff frequency from 1 kHz to 6 kHz. The results of the filtering attack are listed in Table \ref{tab:def_Filter}. It appears that filtering is mostly ineffective against adversarial audios created by \texttt{aaeCAPTCHA} because it does not help improve the SRoA or reduce the WER considerably.

\begin{table}[tp!]
\centering
\begin{tabular}{|c|c|c|c|c|} 
 \hline
 Metric & MP3 & OPUS & AAC & SPEEX \\
 \hline
  WER (\%) & \textbf{99.08} & 105.79 & 107.08 & 104.12 \\
  SRoA (\%) &0.00& 0.00 & 0.00 & 0.00 \\
  SNR (dB) & 13.42 & -3.20 & -3.10 & -3.23 \\
 \hline
\end{tabular}\vspace{-1mm} 
\caption{Robustness of \texttt{aaeCAPTCHA} under Audio Compression attack.} \vspace{-3mm} 
\label{tab:def_compression}
\end{table}

\begin{table*}[tp!]
\centering
\scriptsize
\begin{tabular}{ |C{1.8cm}|C{1.3cm}|C{1.2cm}|C{1cm}|C{1cm}|C{1cm}|C{1cm}|C{1cm}|C{1cm}|C{1cm}|  }
 \hline
 Model & Metric         & No attack   & Quant. ($q=1024$)   & Avg. Smoothing ($k=17$)    &  Med. Smoothing ($k=9$)  & Down-sampling ($5.6$ kHz) & Low-pass Filter ($1.5$ kHz) & Band-pass Filter ($2$ kHz) & Audio Comp. (MP3) \\
 \hline
 \multirow{2}{*}{DS2} & WER (\%) & 78.65 & 80.72  & 77.00 & 76.62  & 85.16 & 87.86 &  83.82 & 73.71  \\
  & SRoA (\%)                    & 0.40  & 0.00   & 0.20  & 0.00   & 0.00  & 0.00  &  0.20 & 0.20  \\  
 \hline
  \multirow{2}{*}{Jasper} & WER (\%)& 56.32 & 31.36 & 56.49 & 50.46 & 58.50 & 54.48 & 54.57 & 51.09  \\
  & SRoA (\%)&                        4.40  & \textbf{16.40} & 4.60  & \textbf{5.80} &  2.20 & 4.00  & 3.60  & \textbf{5.00} \\  
 \hline
  \multirow{2}{*}{W2L+} & WER (\%)& 76.40 & 62.90 & 69.65  & 71.60 & 69.41 & 78.25 & 71.77 & 69.90  \\
  & SRoA (\%)&                       0.40 & \textbf{1.80}  & 0.80   & 0.20  & 0.60  & 0.00  & 0.20  & \textbf{1.20} \\  
 \hline
  \multirow{2}{*}{Lingvo} & WER (\%)& 74.64 & 55.32 & 68.96 & 68.34 & 67.14 & 77.29 & 68.78 & 66.08 \\
  & SRoA (\%)&                      0.40    & \textbf{2.80}  & \textbf{1.60}  & \textbf{1.80}  & 0.80  & 0.60  & 1.00  & \textbf{1.80}\\
 \hline
  \multirow{2}{*}{Kaldi} & WER (\%)& 80.45 & 62.11 & 77.46 & 73.96 & 75.92 & 79.58 & 71.82 & 72.24 \\
  & SRoA (\%)&                       0.40  & \textbf{2.40}  & 0.40  & \textbf{1.40}  & 0.20  & 0.40  & 0.00  & \textbf{ 1.00} \\  
  \hline
  \multirow{2}{*}{GCP STT} & WER (\%)& 34.75  & 26.72 & 35.03 & 41.26 & 40.08 & 41.38 & 36.56 & 51.36 \\
  & SRoA (\%)&                         17.60  & \textbf{23.40} & 17.20 & 11.80 & 11.40 & 12.40 & 17.20 & 4.00 \\  
  \hline
  \multirow{2}{*}{IBM Watson STT} & WER (\%)& 58.74 & 42.21 & 56.25 & 59.65 & 63.57 & 77.14 & 67.46 & 53.77 \\
  & SRoA (\%)&                                5.60  & \textbf{10.80} & 4.80  & 5.00  & 1.80  & 0.60  &  1.20 & \textbf{6.40} \\  
  \hline
  \multirow{2}{*}{Wit.ai STT} & WER (\%)& 36.57 & 31.73 & 36.63 & 42.38 & 40.56 & 40.13 & 39.79 & 36.02 \\
  & SRoA (\%)&                            12.20 & \textbf{17.20} & 12.80 & 11.00 & 9.80  & 11.20 & 11.60 & 11.80 \\  
 \hline
\end{tabular}
\caption{Robustness of \texttt{aaeCAPTCHA} under audio preprocessing attacks evaluated on the tested ASR systems.}
\label{tab:def_eval_on_others}
\end{table*}

\begin{table*}[tp!]
\centering
\scriptsize
\begin{tabular}{ |C{2cm}|C{1.2cm}|C{1.3cm}|C{.7cm}|C{.7cm}|C{.7cm}|C{.7cm}|C{.7cm}|C{1cm}|C{1cm}|C{1cm}| }
 \hline
 Attack  &  Metric       & DeepSpeech  & DS2    & Jasper    &  W2L+    & Lingvo & Kaldi & GCP STT & Watson STT & Wit.ai STT\\
 \hline
 \multirow{2}{*}{Quant. ($q=1024$) } & WER (\%) & 94.85 & 66.59 & 45.41 & 65.74 & 62.42 & 64.62 & 29.38 & 49.79 & 32.68 \\
  & SRoA (\%)                       & 0.00  & 0.40  & 6.00  & 1.80  & 2.60  & 1.40  & 18.40 & 6.20  & 16.40 \\  
 \hline 
\end{tabular}
\caption{Attack performance of ASR systems on adversarial audios crafted using the BPDA algorithm.}
\label{tab:def_break_BDPA}
\end{table*}

\paragraph{Audio compression \cite{das2018adagio, rajaratnam2018speech, zhang2019defending}}
Lossy audio compression techniques have been applied in prior research as attacks against audio adversarial examples. For example, Sch{\"o}nherr \textit{et al.} \cite{schonherr2018adversarial} hypothesized that MP3 compression could be a good countermeasure to adversarial audios since it removes all signals below the human perceptibility threshold. Similarly, other speech compression methods have been studied to mitigate audio adversarial examples \cite{andronic2020mp3, rajaratnam2018speech}. We used MP3, AAC, SPEEX, and OPUS as candidates for the audio compression attack for this experiment.

The WER and SRoA for various audio compression methods are shown in table \ref{tab:def_compression}. We can see that none of these audio compression attack methods is effective against our audio adversarial CAPTCHA system.
Furthermore, the audio compression is largely useless in reducing the WER. However, MP3 compression results in the lowest WER among the four audio compression techniques we studied.

\begin{table*}[tp!]
\centering
\scriptsize
\begin{tabular}{|c|c|c|c|c|c|c|c|c|c|c|} 
 \hline
 Metric & \multicolumn{2}{|c|}{DeepSpeech} & \multicolumn{2}{|c|}{DS2}  & \multicolumn{2}{|c|}{Jasper}  & \multicolumn{2}{|c|}{Lingvo}  & \multicolumn{2}{|c|}{Kaldi} \\ 
   & FGSM AT & PGD AT & FGSM AT & PGD AT & FGSM AT & PGD AT & FGSM AT & PGD AT & FGSM AT & PGD AT \\ 
   \hline
  WER (\%)& 87.11 & 39.28 & 42.60 & 31.52 & 25.77 & 19.61 & 41.38 & 33.76 & 39.95 & 31.12 \\
  SRoA (\%)& 0.20 & 13.20  & 6.80 & 14.60 & 23.60 & 34.80 & 7.80  & 16.60 & 12.80 & 21.00 \\
 \hline
\end{tabular}\vspace{-1mm} 
\caption{Robustness of \texttt{aaeCAPTCHA} under adversarial training attack. \textbf{AT:} Adversarial Training.} \vspace{-2mm}
\label{tab:def_advT}
\end{table*}

\begin{table*}[tp!]
\centering
\scriptsize
\begin{tabular}{|c|c|c|c|c|c|c|c|c|c|c|} 
 \hline
 Metric & \multicolumn{2}{|c|}{DeepSpeech} & \multicolumn{2}{|c|}{DS2}  & \multicolumn{2}{|c|}{Jasper}  & \multicolumn{2}{|c|}{Lingvo}  & \multicolumn{2}{|c|}{Kaldi} \\ 
   & FGSM AT & PGD AT & FGSM AT & PGD AT & FGSM AT & PGD AT & FGSM AT & PGD AT & FGSM AT & PGD AT \\ 
   \hline
  WER (\%)& 86.77 & 100.53 & 46.17 & 45.25 & 34.47 & 37.12 & 49.57  & 46.89 & 53.53 & 52.64 \\
  SRoA (\%)& 0.00 & 0.00   & 5.40  & 6.20  & 14.80 & 13.60 & 4.20   & 6.60 & 5.60  & 6.40  \\
 \hline
\end{tabular}\vspace{-1mm} 
\caption{Attack performance of adversarially trained ASR models on adversarial examples generated against the adversarially trained DeepSpeech model.} \vspace{-3mm}
\label{tab:def_advT_counter}
\end{table*}

\vspace{-1mm}
\subsubsection{Evaluating the impact of audio preprocessing attacks on other state-of-the-art ASR systems}\label{sec:adative_sec_eval:defs_all}~\\
\vspace{-2mm}

In Section \ref{sec:adative_sec_eval:defs_target}, we assessed the performance of audio preprocessing attacks against our target ASR model DeepSpeech. Here, we analyze the attack performance of those methods on both open-source and commercial ASR systems. For this purpose, we first selected the hyperparameter for each attack that resulted in the lowest WER and the highest SRoA during our analysis in Section \ref{sec:adative_sec_eval:defs_target}. Next, we applied each attack separately to the 500 adversarial audios we used for evaluation. Finally, we transcribed these preprocessed adversarial samples using each ASR model.

The results of our experiment are shown in Table \ref{tab:def_eval_on_others}. The quantization attack results in the highest SRoA for most ASR models. For example, quantization helps increase the Jasper ASR model's SRoA from 4.40\% to 16.40\%. Similarly, GCP, IBM Watson, and Wit.ai STTs' success rates of attacks improve under quantization attack. MP3 compression also slightly improves the SRoA for a few ASR models. The other audio preprocessing attacks either degrade the SRoA further or have a negligible impact.

\vspace{-2mm}
\subsubsection{Countering the strongest audio preprocessing attack with adaptive defense}~\\
\vspace{-2mm}

Since quantization with $q=1024$ resulted in the highest success rate of attack for most ASR models, we conducted an experiment to investigate whether this attack against \texttt{aaeCAPTCHA} could be prevented or not. To this end, we incorporated the Backward Pass Differentiable Approximation (BPDA) algorithm into the PGD optimization process to generate adversarial samples resistant to the quantization attack (we encourage the readers to refer to \cite{athalye2018obfuscated} for details). The BPDA and Expectation over Transformation (EOT) \cite{athalye2018synthesizing} have been used in previous studies to break attacks against adversarial examples \cite{NEURIPS2020_tramer_onadaptiveattacks}.
Table \ref{tab:def_break_BDPA} lists the results of our experiment. We can see that adversarial audio generated with BPDA reduces the SRoA for all ASR models significantly.
For example, the Jasper model's SRoA is more than 16\% under the quantization ($q=1024$) attack, whereas it is only 6\% for BPDA-generated adversarial audios under the same attack.
However, BPDA degrades the SNR slightly by adding more distortion to adversarial audios.
The mean SNR and $||\delta||_1$ for BPDA-crafted adversarial audios are 15.58 dB and 271.20, respectively.
In comparison, the original mean SNR and $||\delta||_1$ are 15.89 dB and 256.61, respectively.
To conclude, our findings indicate that the audio preprocessing attacks against \texttt{aaeCAPTCHA} could be nullified by crafting adversarial audio with BPDA and other adaptive defenses from adversarial machine learning research.


\subsubsection{Adversarial training ~\cite{sun2018training, sun2019adversarial, wang2019adversarial} attack against aaeCAPTCHA}\label{sec:adative_sec_eval:AT}~\\
\vspace{-2mm}

To break \texttt{aaeCAPTCHA}, an adaptive adversary can train a local ASR model specifically on adversarial audio samples generated by our system. Such a training process is known as ``adversarial training'' \cite{szegedy2013intriguing} and is regarded as one of the most effective countermeasures to make neural networks more robust to adversarial samples. Adversarial training improves the model's robustness to adversarial attacks by augmenting training data with adversarial examples \cite{goodfellow2014explaining, huang2015learning}. Efficient adversarial attacks like FGSM, I-FGSM, and PGD are commonly employed for adversarial training \cite{madry2017towards, kurakin2016adversarial, tramer2017ensemble}. To investigate the robustness of \texttt{aaeCAPTCHA} against the adversarial training attack, we evaluated the attack performance of five different adversarially trained local ASR models. The initial models were pretrained on the LibriSpeech dataset.

\paragraph{Evaluation setup and methodology} We employed both PGD and FGSM for performing the adversarial training attack. For PGD adversarial training, we selected the same hyperparameters we used for generating our audio adversarial CAPTCHAs. Specifically, we chose $\epsilon=350$, $steps=50$, and $\alpha=40$. We selected those particular hyperparameters because we assumed a powerful attacker who has complete knowledge of our audio adversarial CAPTCHA generation algorithm. For FGSM adversarial training, we set $\epsilon=350$.

We used all the audio samples with a duration of less than 16 seconds from the LibriSpeech \texttt{train-clean-100} dataset for training. We also used the whole \texttt{dev-other} dataset and \texttt{dev-clean} dataset. However, we excluded the 500 audios we used throughout this paper for evaluation from the \texttt{dev-clean} dataset.

For adversarially training the ASR models, we adopted the adversarial training strategy from prior research \cite{abdullah2019hear}. Specifically, we first perturbed all three datasets described above using the PGD algorithm. Perturbed samples were computed against our target ASR model DeepSpeech. Next, we trained DeepSpeech, DeepSpeech 2, Jasper, Lingvo, and Kaldi on PGD-generated perturbed samples for 32 epochs with a batch size of 64. However, our evaluations were conducted on the checkpoints for which we obtained the lowest validation loss on the corresponding validation sets. We followed the same method for FGSM adversarial training. Adversarially trained ASR models' attack performances were evaluated on 500 adversarial audios crafted using the PGD (with $\epsilon=350$, $steps=50$, and $\alpha=40$) algorithm against the DeepSpeech ASR model.

\paragraph{Results} The experimental results of the adversarial training attack are presented in Table \ref{tab:def_advT}. PGD adversarial training improves the success rate of attack for all ASR models. The Jasper ASR model obtains the highest SRoA. FGSM adversarial training increases the SRoA for these models as well. Overall, the adversarial training attack is more effective than audio preprocessing attacks against \texttt{aaeCAPTCHA}.

Next, we studied whether the adversarial training attack could be countered. A straightforward approach is to craft adversarial examples against our adversarially trained target network. To test the efficacy of this method, we generated 500 audio adversarial examples using the PGD algorithm with $\epsilon=350$, $steps=50$, and $\alpha=40$ against our adversarially trained DeepSpeech ASR model. We then tested the attack performance of other adversarially trained ASR models on these audios. Table \ref{tab:def_advT_counter} depicts the results. The SRoA for both FGSM and PGD adversarially trained DeepSpeech models falls to 0\%. 
Except for the Jasper, the SRoA remains below 7\% for all other adversarially trained ASR models.
However, this defense method against the adversarial training attack causes some minor distortion in the updated adversarial audios. Specifically, the mean SNR and $||\delta||_1$ for updated adversarial audios are 15.73 dB and 263.18, while the mean SNR and $||\delta||_1$ for initial adversarial audios are 15.89 dB and 256.61. Moreover, the continuous adversarial training of the target ASR model and the regeneration of adversarial audios are likely to be less viable in practice.



\begin{table}[tp!]
\centering
\scriptsize
\begin{tabular}{|C{1.5cm}|C{.7cm}|C{.5cm}|C{.5cm}|C{.5cm}|C{.8cm}|C{.5cm}|} 
 \hline
                    & Normal & $\epsilon:250$ &  $\epsilon:300$ & $\epsilon:350$ & $\epsilon:350$ (BPDA) &  $\epsilon:400$  \\ 
  \hline
  Total             & 195    &  197           &  194            &  198           & 197                   & 196           \\
  \hline                   
  Success rate (\%) & 85.13  & 76.65 & 72.68 & 74.24 & 72.08 & 70.41\\
  \hline
  Average time (s)  & 41.83 & 43.76 &  44.83 & 48.86 & 45.11 & 59.69 \\
  \hline
  Median time (s)   & 28.50  & 27.50  &  29.00 & 28.00   & 31.00  & 48.00  \\
  \hline
  Std time (s)      & 28.74 & 37.25 &  35.99 & 46.81 & 39.58 & 42.11\\
 \hline
 
\end{tabular}
\caption{Human success rate and completion time for audio CAPTCHAs with varying perturbation budgets.}
\label{tab:user_study:sec_usability_tradeoffs}
\end{table}

\begin{table*}[tp!]
\centering
\scriptsize
\begin{tabular}{|C{1cm}|C{.55cm}|C{.55cm}|C{.55cm}|C{.55cm}|C{.55cm}|C{.55cm}|C{.55cm}|C{.55cm}|C{.55cm}|C{.55cm}|C{.55cm}|C{.55cm}|C{.5cm}|C{.55cm}|C{.5cm}|C{.5cm}|C{.5cm}|} 
 \hline
    & \multicolumn{2}{c|}{Overall} & \multicolumn{14}{c|}{Transcription length (word)} \\
\hline
                 & & &  \multicolumn{2}{c|}{6}  &  \multicolumn{2}{c|}{7}  &  \multicolumn{2}{c|}{8}  &  \multicolumn{2}{c|}{9}  &  \multicolumn{2}{c|}{10}  & \multicolumn{2}{c|}{ 11}  &  \multicolumn{2}{c|}{12}  \\ 
                     & N  & A  & N  & A  & N  & A   & N  & A  & N  & A  & N  & A & N & A & N & A   \\                   
 \hline
   Total             & 195     &   198   & 37 &  32  & 25 &  38 & 39 &   29  & 33 &  34  & 23 & 23  & 22 &  18  & 16  & 24 \\
   \hline
   Success rate (\%) &   85.13  &  74.24  & 91.89 &  78.12 & 84.00 & 86.84 & 87.18 & 82.76& 90.91 &  61.76& 82.61 & 73.91& 72.73 & 72.22 & 75.00  & 58.33 \\
  \hline
    Average time (s) &   41.83   &  48.86 & 39.64 & 54.06 & 53.14 & 44.00 & 49.24 & 55.10 & 32.53 & 33.07 & 39.47 & 62.23 & 41.35 & 41.36 & 39.88 & 48.09 \\
    \hline
    Median time (s)  &   28.50  &  28.00  & 24.50 &  48.00 & 22.00 &25.00 & 42.00 & 31.00 & 23.00 &  22.00 & 25.00 & 26.00 & 31.00 &  28.00 & 27.00 & 28.00\\
    \hline
    Std time (s)     &   28.74   &  46.81  & 32.68 & 54.21  & 51.06 & 43.41 & 26.39 & 52.13 & 21.89 & 31.13 & 29.21 & 59.63 & 24.04 & 36.16 & 25.15 & 43.33\\
 \hline
 
\end{tabular}
\caption{Human success rate and completion time for normal and adversarial ($\epsilon=350$) CAPTCHAs with different transcription lengths. \textbf{N:} Normal Audios (Baseline), \textbf{A:} Adversarial Audios.}
\vspace{-3mm} 
\label{tab:user_study:normal_vs_adv}
\end{table*}



\vspace{-1mm}
\section{Usability Evaluation} \label{sec:usability} \vspace{-1mm} 
\vspace{-1mm}

We comprehensively investigated the security of \texttt{aaeCAPTCHA} under both general and adaptive security settings in Section \ref{sec:sec_eval}. In this section, we evaluate the usability of the \texttt{aaeCAPTCHA} proof-of-concept implementation. We recruited 100 users from the Amazon Mechanical Turk (MTurk) crowdsourcing marketplace for this study. Participants were from the United States and had a Human Intelligence Task (HIT) approval rate greater than or equal to 85\%.

\paragraph{Evaluation setup} We generated audio adversarial CAPTCHAs with various perturbation budgets using the PGD algorithm against the DeepSpeech ASR model. In particular, we used 4 perturbation budgets for this experiment: $\epsilon=250, 300, 350, 400$. We followed the same approach as discussed in Section \ref{sec:pgd_opt_hp} for selecting the number of PGD steps $steps$, and learning rate $\alpha$ for each chosen $\epsilon$. 
The details of the evaluation setup are reported in Table \ref{tab:user_study:setup} in Appendix \ref{sec:appndx}.
For $\epsilon=350$, we also crafted adversarial audios with BPDA. We generated 500 adversarial audios for each setting using the same dataset (LibriSpeech \texttt{dev-clean}) we used for security evaluation. We set the baseline as the usability of normal audios.

\paragraph{Methodology} Participants were presented with 12 audio CAPTCHAs, including two normal audios, two adversarial audios each from the four perturbation budgets, and two adversarial audios crafted with BPDA. The audio CAPTCHAs for each corresponding category were selected randomly for each participant. Finally, our system asked participants to transcribe the audios.
One hundred users submitted a total of 1,200 solutions (transcriptions). We discarded 23 of them, for which the participants took an unusually long time to submit the transcription. For example, some users took more than 350 seconds to solve a 2-second audio CAPTCHA. Therefore, our evaluation was conducted on the remaining 1,177 audios. For computing the success rate, we consider an audio CAPTCHA to be successfully solved if the WER between the submitted transcription and the ground transcription is zero.

\paragraph{Ethics} Our usability evaluation involved human subjects. In our study, we did not collect or store any personally identifiable information about Amazon MTurk workers \footnote{Unfortunately, for this reason, we were unable to report more detailed results (\textit{e.g.}, usability across different age groups, usability vs. gender, etc.).}. There was no risk to participating in the study other than the risks associated with everyday computer use.
Therefore, our Institutional Review Board (IRB) application was approved (Category: Exempt).  

\vspace{-2mm}
\subsection{Results and Analysis}
\vspace{-3mm}

Table \ref{tab:user_study:sec_usability_tradeoffs} shows the success rates and completion times for normal audio CAPTCHAs as well as adversarial CAPTCHAs with varying perturbation budgets. As expected, the highest success rate is obtained for normal audios. The success rate usually decreases when the perturbation budget increases. Compared to the baseline, adversarial audios created with $\epsilon = 350$ have about an 11\% lower success rate. Adversarial audio crafted with BPDA further degrades the success rate slightly. Adversarial audios generated by $\epsilon=400$ have the lowest success rate. Table \ref{tab:user_study:normal_vs_adv} also lists the completion times for successfully solved CAPTCHAs. The average completion time increases with the increase in the perturbation budget. Generally, adversarial audios with a higher perturbation budget $\epsilon$ are more robust against ASR model attacks. However, this negatively impacts usability. The attack performances of ASR systems on adversarial audios with the four perturbation budgets are listed in Tables \ref{tab:machine:pgd_200_40_60},  \ref{tab:tab:machine:pgd_300_30_70}, \ref{tab:transf_opnsrc}, \ref{tab:transf_commer_stt}, and \ref{tab:machine:pgd_400_40_50}.
Ultimately, the CAPTCHA developer will need to choose the optimal perturbation budget $\epsilon$ along with other hyperparameters, considering the trade-off between security and usability for deploying audio adversarial CAPTCHA systems.

Table \ref{tab:user_study:normal_vs_adv} shows the success rates of normal and adversarial (with $\epsilon=350$) audios with varying transcription lengths. The general trend is that participants are more successful at solving audio CAPTCHAs with shorter transcription lengths. However, we can notice some discrepancies in completion times. By analyzing the completion times, we found that some participants took an abnormally long time (\textit{e.g.}, over 120 seconds) to transcribe audios with very short transcription lengths (\textit{e.g.}, six words). We suspect the participants were mostly inactive during that period.

Based on usability evaluation results, we finally conclude that \texttt{aaeCAPTCHA} achieves high robustness at a moderate usability cost compared to normal audio CAPTCHAs.

\vspace{-1mm} 


\section{Discussion}

Diverting from traditional designs, we introduced \texttt{aaeCAPTCHA}, a new audio CAPTCHA system that uses adversarial audios as CAPTCHAs to safeguard the system against attacks using ASR models. The \texttt{aaeCAPTCHA} design exploits the vulnerability of ASR models to adversarial examples and uses it to defend the CAPTCHA system against them. We discussed why prior audio adversarial generation methods are not suitable for audio CAPTCHA designs. Through extensive experimentation, we showed that the PGD algorithm with a higher maximum allowed perturbation $\epsilon$ could be utilized to generate audio adversarial CAPTCHAs. We validated the robustness of \texttt{aaeCAPTCHA} against ASR attacks by conducting a rigorous security evaluation employing five state-of-the-art ASR systems and three commercial speech-to-text services.
Furthermore, we conducted a usability analysis of the \texttt{aaeCAPTCHA} scheme and showed the trade-off between robustness and usability. It is worth noting that \texttt{aaeCAPTCHA} is not designed to replace traditional audio CAPTCHAs. Instead, it is intended as an enhancement to existing audio CAPTCHA systems. Our overall findings highlight that \texttt{aaeCAPTCHA} can significantly improve the robustness of traditional audio CAPTCHAs without greatly sacrificing usability.

\vspace{-2mm}
\subsection{Limitations and Future Work}
\vspace{-1mm}

We believe \texttt{aaeCAPTCHA} can be improved from many perspectives as an effort to develop robust audio adversarial CAPTCHA systems. The limitations of this work, as well as future work, are discussed below. 

First, as previously stated, \texttt{aaeCAPTCHA} anticipates human-intelligible rather than human-imperceptible audio adversarial perturbations. However, we established the human-intelligible perturbation budgets based on our experience and preliminary evaluation in our experiments. In other words, we do not yet have a standard for quantifying human-intelligible perturbation. As a result, more dedicated research into understanding and measuring the trade-off between CAPTCHA security and usability is expected. 

Second, to defend against the adversarial training attack, we proposed periodic retraining of the target ASR model on previously generated adversarial samples. This recommendation is based on a powerful adversary having complete knowledge of the CAPTCHA system. As such, the adversarial training attack in Section \ref{sec:adative_sec_eval:AT} demonstrates a theoretical bound on the robustness of \texttt{aaeCAPTCHA}. However, the attacker is likely to have no or limited knowledge of our CAPTCHA generation algorithm in a real-world setting, and the above measure might not be required. Therefore, further research is required to investigate a more practical adversarial training attack against \texttt{aaeCAPTCHA} and defenses against such an attack.

Third, our current approach does not focus on minimizing the perceptibility of adversarial perturbations. Prior research shows that by employing psychoacoustic hiding techniques, it is possible to generate adversarial audio with less audible distortion \cite{schonherr2018adversarial, abdullah2021beyond}. Incorporating such a method in our audio adversarial CAPTCHA generator might make it possible to create audio adversarial CAPTCHAs that are more human-intelligible and whose distortions are less perceptible.

Fourth, we attacked the acoustic model of our target ASR system while crafting adversarial CAPTCHAs. While we achieved high adversarial success rates using this method against various ASR models we evaluated, another interesting direction to pursue is directly attacking the audio processing pipeline stages (\textit{e.g.}, feature extraction module) before the model, which are quite similar across modern ASR systems.
Such a technique is independent of a specific DNN acoustic model and should result in adversarial audios that are efficient, less distorted, and transferable. Previous research has shown that such a strategy is feasible \cite{abdullah2019hear}.
However, more studies are required to verify the viability of such techniques for audio CAPTCHA generation purposes.

\vspace{-1mm}
\section{Conclusion} 
\vspace{-1mm}




In this paper, we introduce the design and implementation of an audio adversarial CAPTCHA system called \texttt{aaeCAPTCHA}. The \texttt{aaeCAPTCHA} system exploits audio adversarial examples as audio CAPTCHA challenges. In addition, we conducted an extensive security evaluation of our new CAPTCHA design. Our experimental results show that \texttt{aaeCAPTCHA} is highly secure against automated attacks using state-of-the-art speech recognition technologies.
Furthermore, we conducted a user study to investigate the usability of the \texttt{aaeCAPTCHA} proof-of-concept implementation, and our findings show that the system maintains good usability.
Finally, our comprehensive evaluation results demonstrate that \texttt{aaeCAPTCHA} can significantly enhance the security of traditional audio CAPTCHA schemes.

\vspace{-2mm}
\section*{Acknowledgments} 
\vspace{-2mm}
The authors thank the anonymous reviewers for their valuable comments that improved this paper.
This work is supported in part by the US NSF under grants OIA-1946231 and CNS-2117785.

\bibliographystyle{plain}
\bibliography{reference}

\appendices
\section{}\label{sec:appndx}

\begin{table}[tp!]
\centering
\scriptsize
\begin{tabular}{|C{1cm}|C{1cm}|C{1cm}|C{1cm}|C{1cm}|C{1cm}|} 
 \hline
                 & $\epsilon:250$ &  $\epsilon:300$ & $\epsilon:350$ & $\epsilon:350$ (BPDA) &$\epsilon:400$  \\ 
  \hline
  Parameters     & $steps:40$, $\alpha:60$ & $steps:30$, $\alpha:70$ &  $steps:50$, $\alpha:40$ & $steps:50$, $\alpha:40$ & $steps:40$, $\alpha:50$ \\
  \hline
  WER (\%)       & 111.93 & 111.53 & 111.88 & 94.85 & 109.90 \\
  \hline
  SRoA (\%)      & 0.00   & 0.00   & 0.00   & 0.00  & 0.00 \\
  \hline
  SNR (dB)       & 17.71  & 16.51  & 15.89  & 15.58 & 14.87 \\
 \hline
  $||\delta||_1$ & 219.42 & 247.06 & 256.61 & 271.20 & 286.76 \\
  \hline
\end{tabular}
\caption{The selected perturbation budget $\epsilon$ values and other hyperparameters for the PGD algorithm used in the usability analysis. Adversarial audios were generated and evaluated against the DeepSpeech ASR model.}  
\label{tab:user_study:setup}
\end{table}

\begin{table*}[htp!]
    \centering
    \begin{tabular}{|C{3cm}|C{5cm}|C{5cm}|}
    \hline
      Model & Normal  & Adversarial \\
       \hline
      DS2   & yes he's mere curier in all his movements & t the alase husing off clolier off the can almasy lafrest \\
       \hline
      Jasper & yes he's mercurial in all his movements & but the earth osing off clonerod an alms lead for ust \\
       \hline
      W2L+ &  yes he's mercurial in all his movements & but the earths oozing of clonier of in owms live for ust \\
       \hline
      Lingvo & yes he's more cruel in all his movements & for the honest who was in our flonieron than could arms he had for us \\
       \hline
      Kaldi & yes he's mercurial in all his movements & for the erse who using off claudia yer ole kit arms he lay four hissed \\
       \hline
      GCP STT & yes he's mercurial in all his movements & she's not coming out and all his new friends \\
       \hline 
      IBM Watson STT & yes he's mercurial in all his movements & yes i can only \\
       \hline
      Wit.ai STT & yes he's mercurial in all his movements & yes using our career it always made for us \\
       \hline 
       \multicolumn{3}{l}{\small *Ground truth transcription is ``yes he's mercurial in all his movements''.} 
    \end{tabular}
    \caption{Transcriptions returned by various ASR systems for a sample audio file and its adversarial version. The results for GCP, IBM Watson, and Wit.ai STTs were obtained in January 2022.}
    \label{tab:asr_out_sample}
\end{table*}


\begin{table*}[htp!]
\centering
\scriptsize
\begin{tabular}{ |C{.8cm}|C{1.3cm}|C{.8cm}|C{.8cm}|C{.8cm}|C{.8cm}|C{.8cm}|C{.8cm}|C{1.2cm}|C{.8cm}| }
 \hline
    Metric        & DeepSpeech & DS2 & Jasper & W2L+ & Lingvo & Kaldi & GCP STT & IBM Watson STT & Wit.ai STT \\
  \hline
  WER (\%)  & 111.93  & 69.96 & 46.61 & 68.60 & 65.89 & 98.55 & 28.56 & 49.49 & 29.38 \\
  \hline
  SRoA (\%) &  0.00 & 0.80  & 8.60  & 1.40  & 1.80  & 0.00 & 23.60 & 7.00 & 20.80 \\
 \hline
\end{tabular}
\caption{Attack performance of ASR systems on adversarial audios crafted using the PGD algorithm with $\epsilon=250$, $steps=40$, and $\alpha=60$.}
\label{tab:machine:pgd_200_40_60}
\end{table*}

\begin{table*}[htp!]
\centering
\scriptsize
\begin{tabular}{ |C{.8cm}|C{1.3cm}|C{.8cm}|C{.8cm}|C{.8cm}|C{.8cm}|C{.8cm}|C{.8cm}|C{1.2cm}|C{.8cm}| }
 \hline
    Metric        & DeepSpeech  & DS2 & Jasper & W2L+ & Lingvo & Kaldi & GCP STT & IBM Watson STT & Wit.ai STT \\
  \hline
  WER (\%)  & 111.53  & 75.60 & 52.33 & 74.55 & 69.41 & 98.56 & 32.32  & 54.74 & 33.22 \\
  \hline
  SRoA (\%) &  0.00  & 0.60 & 5.60 & 0.60 & 2.00 & 0.00  & 21.60 & 5.20 & 17.60 \\
 \hline
\end{tabular}
\caption{Attack performance of ASR systems on adversarial audios crafted using the PGD algorithm with $\epsilon=300$, $steps=30$, and $\alpha=70$.}
\label{tab:tab:machine:pgd_300_30_70}
\end{table*}

\begin{table*}[htp!]
\centering
\begin{tabular}{ |C{.8cm}|C{1.3cm}|C{.8cm}|C{.8cm}|C{.8cm}|C{.8cm}|C{.8cm}|C{.8cm}|C{1.2cm}|C{.8cm}| }
 \hline
    Metric        & DeepSpeech  & DS2 & Jasper & W2L+ & Lingvo & Kaldi & GCP STT & IBM Watson STT & Wit.ai STT \\
  \hline
  WER (\%)  & 109.90  & 80.35 & 59.99 & 80.23 & 77.44 & 98.22 & 35.49  & 60.89 & 39.07  \\
  \hline
  SRoA (\%) & 0.00    & 0.00 & 3.40 & 0.20 & 0.60 & 0.00 & 18.00 & 2.60 & 11.00 \\
 \hline
\end{tabular}
\caption{Attack performance of ASR systems on adversarial audios crafted using the PGD algorithm with $\epsilon=400$, $steps=40$, and $\alpha=50$.}
\label{tab:machine:pgd_400_40_50}
\end{table*}


\end{document}